\documentclass[letterpaper,11pt]{JHEP3}
\usepackage{amsmath,amssymb}
\usepackage{latexsym}
\usepackage{epsfig}
\usepackage{cite}
\usepackage[english]{babel}

\usepackage{graphicx}
\usepackage{epsf}
\usepackage{amsfonts}
\usepackage{epic}
\usepackage{amsthm,multibox}
\usepackage{amsbsy}

\newcommand{\eq}{\begin{equation}}
\newcommand{\feq}{\end{equation}}
\newcommand{\eqn}{\begin{eqnarray}}
\newcommand{\feqn}{\end{eqnarray}}
\newcommand{\arr}{\begin{eqnarray*}}
\newcommand{\farr}{\end{eqnarray*}}

\font\mybb=msbm10 at 12pt
\def\bb#1{\hbox{\mybb#1}}

\def\bR {\bb{R}}

\def\sls#1{\setbox0=\hbox{$#1$}
   \dimen0=\wd0
   \setbox1=\hbox{/} \dimen1=\wd1
   \ifdim\dimen0>\dimen1
      \rlap{\hbox to \dimen0{\hfil/\hfil}}
      #1
   \else
      \rlap{\hbox to \dimen1{\hfil$#1$\hfil}}
      /
   \fi}

\title{Generalized instantons in ${\cal N}=4$ super Yang-Mills theory
and spinorial geometry}

\author{St\'ephane Detournay,$^{abc}$ Dietmar Klemm,$^{bc}$ and
Carlo Pedroli$^b$ \\
$^a$ Kavli Institute for Theoretical Physics, \\
\hspace*{0.15cm} University of California, \\
\hspace*{0.15cm} Santa Barbara, CA 93106, USA. \\
$^b$ Dipartimento di Fisica dell'Universit\`a di Milano, \\
\hspace*{0.15cm} Via Celoria 16, I-20133 Milano. \\
$^c$ INFN, Sezione di Milano, Via Celoria 16, I-20133 Milano. \\
}
\preprint{IFUM-944-FT\\NSF-KITP-09-128}

\abstract{Using spinorial geometry techniques, we classify the supersymmetric
solutions of euclidean ${\cal N}=4$ super Yang-Mills theory. These backgrounds
represent generalizations of instantons with nontrivial scalar fields turned
on, and satisfy some constraints that bear a similarity with the Hitchin
equations, and contain the Donaldson equations as a special subcase.
It turns out that these constraints can be obtained by dimensional reduction of
the octonionic instanton equations, and may be rephrased in terms of a
selfduality-like condition for a complex connection.
We also show that the supersymmetry conditions imply the equations of motion
only partially.}

\keywords{AdS/CFT correspondence, supersymmetric gauge theory, instantons}

\begin{document}

\section{Introduction}
\label{intro}

Supersymmetric backgrounds play an important role both in supergravity
and in supersymmetric gauge theories. In the former, they include
for instance BPS black holes, whose study has given many hints on the
quantum nature of spacetime (see e.g. \cite{Balasubramanian:2005mg}). In the latter, there are e.g.~monopoles or
instantons among the supersymmetric solutions. Instantons are also
important in nonsupersymmetric field theories, like in QCD, where the
nonperturbative chiral $U(1)$ anomaly in an instanton background leads to
baryon number violation and to a solution of the $U(1)$
problem \cite{'tHooft:1976fv,'tHooft:1976up}. Viewed as a solution of
supersymmetric gauge theories, the Yang-Mills instanton \cite{Belavin:1975fg}
preserves half of the supersymmetries, and has been important in checks
of the AdS/CFT correspondence beyond the perturbative level,
cf.~\cite{Bianchi:2007ft} for a review.

In view of this, it is desirable to dispose of a complete classification
of supersymmetric backgrounds of super Yang-Mills theories. This paper
represents a first step in this direction for euclidean ${\cal N}=4$ super
Yang-Mills theory, which is the most important one in the context of the
AdS/CFT correspondence. We shall directly solve the Killing spinor equations
using spinorial geometry techniques, that have been successfully applied
in the past in classifying supergravity solutions \cite{Gillard:2004xq}.
The basic ingredients are an oscillator basis for the spinors in terms of
forms and the use of the symmetries to transform them to a preferred
representative of their orbit. In this way one can construct a linear
system for the background fields from any (set of) Killing spinor(s).
It will turn out that this linear system describes generalizations of
instantons that include also nonvanishing scalars. Quite remarkably, this
system can be obtained by dimensional reduction of the octonionic instanton
equations in eight dimensions.

The remainder of this paper is organized as follows: In the next section,
we introduce euclidean ${\cal N}=4$ super Yang-Mills theory and, to make
this paper self-contained, briefly discuss the usual instanton solutions.
In Sect. \ref{spinorial}, the essential information needed
to realize spinors in terms of forms is summarized. Section \ref{Classification}
represents the main part of this work, in which we determine first the
Killing spinor representatives as well as their stability subgroup,
and subsequently obtain the linear system for the background fields,
which is then discussed and related to the octonionic instanton equations.
After that, we impose more Killing spinors and investigate which fractions
of supersymmetry are possible.
In Section \ref{susyeom} we discuss to what extent the supersymmetry
conditions imply the equations of motion. We conclude in \ref{final} with
some final remarks. An appendix contains our notations and conventions.

\section{Euclidean ${\cal N}=4$ SYM theory and instantons}\label{SYM}

The Lagrangian of euclidean ${\cal N}=4$ super Yang-Mills (SYM) theory in four
dimensions is \cite{Brink:1976bc,Vandoren:2008xg}
\begin{eqnarray}
    \mathcal{L}&=&\frac1{g^2}\mbox{tr}\Big\{\frac{1}{2}F_{\mu\nu}F^{\mu\nu}
    -i\bar\lambda^{\dot\alpha}_A\sls{\bar{D}}_{\dot\alpha\beta}\lambda^{\beta,A}
    -i\lambda_{\alpha}^A\sls{D}^{\alpha\dot\beta}\bar\lambda_{\dot\beta,A}+
    \frac12(D_\mu\bar\phi_{AB})(D^\mu\phi^{AB})\nonumber \\
    &&-\sqrt2\bar\phi_{AB}\{\lambda^{\alpha,A},\lambda_\alpha^B\}
    -\sqrt2\phi^{AB}\{\bar\lambda^{\dot\alpha}_A,\bar\lambda_{\dot\alpha,B}\}
    +\frac18[\phi^{AB},\phi^{CD}][\bar\phi_{AB},\bar\phi_{CD}]
    \Big\}\,. \label{eq: lagrangiana euclidea}
\end{eqnarray}
It can be obtained by dimensionally reducing ${\cal N}=1$ SYM in
10-dimensional Minkowski space-time on a six-torus with one time and five space
coordinates \cite{Vandoren:2008xg,Blau:1997pp,Belitsky:2000ii}\footnote{Note
that one cannot get the theory \eqref{eq: lagrangiana euclidea} by
simply Wick-rotating Minkowskian
${\cal N}=4$ SYM. In fact, the Wick rotation of the bosonic sector of
${\cal N}=4$ SYM theory has no supersymmetric completion. See \cite{Hull:1998vg}
for a discussion. Notice also that in \eqref{eq: lagrangiana euclidea}, the
scalar coming from the time component $A_0$ of the ten-dimensional vector
potential has a kinetic term of the wrong sign, so the theory has ghosts.
It is easy to see that this (real) scalar is $i(\phi^{14}+\phi^{23})$.}. The
ten-dimensional Lorentz group $SO(9,1)$ reduces then to $SO(4)\times SO(5,1)$,
with compact space-time group $SO(4)$ and non-compact R-symmetry $SO(5,1)$.
The fields present are the gauge field $A_\mu$ with
field strength $F_{\mu \nu}$, Spin$(4)$ Weyl spinors $\lambda^{\alpha,A}$ and
$\bar\lambda_{\dot\alpha,A}$, respectively right- and left-handed, and
six scalars $\phi^{AB}$ antisymmetric in the R-symmetry group indices
$A,B=1,\ldots,4$. Their duals are defined as
$\bar\phi_{AB}=\frac12\epsilon_{ABCD}\phi^{CD}$. The trace is taken in the adjoint
representation of the gauge group to which all the fields belong (the gauge group
indices are omitted for clarity). The Majorana-Weyl condition imposed on the
ten-dimensional spinors implies the so-called symplectic Majorana condition on the
four-dimensional Weyl spinors\footnote{Unless specified otherwise, equations which
involve complex conjugation of fields will be understood as not Lie algebra
valued, i.e.~they hold for the components $\lambda^{a,\alpha,A}$ etc.} (the usual
Majorana condition cannot be imposed in four euclidean dimensions):
\begin{equation}\label{eq: simplettica}
  \begin{split}
    &(\lambda^{\alpha,A})^*=-\eta_{1AB}\epsilon_{\alpha\beta}\lambda^{\beta,B}=\eta_{1AB}\lambda^B_\alpha\\
    &(\bar\lambda_{\dot\alpha,A})^*=-\eta_1^{AB}\epsilon^{\dot\alpha\dot\beta}\bar\lambda_{\dot\beta,B}=-\eta_1^{AB}\bar\lambda_B^{\dot\alpha}\,,
  \end{split}
\end{equation}
where the definition of the 't Hooft symbols $\eta_{aAB}$ and other conventions
are collected in appendix \ref{conv}. The scalar fields on the other hand are
constrained by the reality condition
\begin{equation}\label{ScalarsCC}
  (\phi^{AB})^*=\eta_{1AC}\phi^{CD}\eta_{1DB}\,,
\end{equation}
or explicitly
\begin{eqnarray}
(\phi^{12})^* = \phi^{34} \; ,\;(\phi^{13})^* = -\phi^{24} \;,\;(\phi^{14})^* = -\phi^{14} \;,\;(\phi^{23})^* = -\phi^{23} \;,
\end{eqnarray}
since they originate from the dimensional reduction of the real 10-dimensional gauge field.

The Lagrangian \eqref{eq: lagrangiana euclidea} is invariant under the
supersymmetry transformations
\begin{eqnarray}\label{SusyEqCompl}
\delta A_\mu&=&-i\bar\xi_A^{\dot\alpha}\bar\sigma_{\mu\dot\alpha\beta}
               \lambda^{\beta,A}+i\bar\lambda_{\dot\beta,A}
               \sigma_{\mu}^{\alpha\dot\beta}\xi_{\alpha}^A\ , \nonumber \\
\delta\phi^{AB}&=&\sqrt2\left(\xi^{\alpha,A}\lambda^B_\alpha-\xi^{\alpha,B}
                 \lambda^A_\alpha+\epsilon^{ABCD}\bar\xi^{\dot\alpha}_C
                 \bar\lambda_{\dot\alpha,D}\right)\ , \nonumber \\
\delta\lambda^{\alpha,A}&=&-\frac12\sigma^{\mu\nu\alpha}_{\ \ \ \ \beta}
                          F_{\mu\nu}\xi^{\beta,A}-i\sqrt2\bar{\xi}_{\dot\alpha,B}
                          \sls{D}^{\alpha\dot\alpha}\phi^{AB}+[\phi^{AB},
                          \bar\phi_{BC}]\xi^{\alpha,C}\ , \nonumber \\
\delta\bar\lambda_{\dot\alpha,A}&=&-\frac12\bar\sigma^{\mu\nu\ \dot\beta}_{\ \ \dot
                                  \alpha}F_{\mu\nu}\bar\xi_{\dot\beta,A}+i\sqrt2
                                  \xi^{\alpha,B}\sls{D}_{\dot\alpha\alpha}\bar
                                  \phi_{AB}+[\bar\phi_{AB},\phi^{BC}]\bar
                                  \xi_{\dot\alpha,C}\ ,
\end{eqnarray}
whose fermionic parameters $\xi^A$ and $\bar{\xi}_A$ themselves have to
satisfy the symplectic Majorana condition \eqref{eq: simplettica}. The equations
of motion derived from \eqref{eq: lagrangiana euclidea} read
\begin{eqnarray}
&&D^\nu F_{\nu\mu}-i\{\bar\lambda^{\dot\alpha}_A\bar\sigma_{\mu\dot\alpha\beta},
  \lambda^{\beta,A}\}-\frac12[\bar\phi_{AB},D_{\mu}\phi^{AB}]=0\ , \label{eomF} \\
&&D^2\phi^{AB}+\sqrt2\{\lambda^{\alpha,A},\lambda^B_{\alpha}\}+\frac1{\sqrt2}
  \epsilon^{ABCD}\{\bar\lambda^{\dot\alpha}_C,\bar\lambda_{\dot\alpha,D}\}
  -\frac12[\bar\phi_{CD},[\phi^{AB},\phi^{CD}]]=0\ , \label{eomphi} \\
&&\sls{\bar{D}}_{\dot\alpha\beta}\lambda^{\beta,A}+i\sqrt2[\phi^{AB},
  \bar\lambda_{\dot\alpha,B}]=0\ , \label{eomlambda} \\
&&\sls{D}^{\alpha\dot\beta}\bar\lambda_{\dot\beta,A}-i\sqrt2[\bar\phi_{AB},
  \lambda^{\alpha,B}]=0\ . \label{eombarlambda}
\end{eqnarray}

A notable solution of the pure euclidean Yang-Mills field equations is given
by a field strength which is either selfdual or anti-selfdual,
\begin{equation}\label{instanton}
F_{\mu\nu} = \pm\frac12\epsilon_{\mu\nu\rho\sigma}F^{\rho\sigma}\ .
\end{equation}
These solutions correspond to (anti-)instantons, i.e. finite-action solutions
to the euclidean theory. In a given topological sector (characterized by the
instanton or winding number $k$), the solutions \eqref{instanton} actually
minimize the action (for reviews on instantons, see
e.g.~\cite{Vandoren:2008xg,Belitsky:2000ws,Dorey:2002ik,Bianchi:2007ft}).
These configurations also generically correspond to solutions preserving some
fraction of the supersymmetries in SYM theories. In the ${\cal N}=2$ case for
instance (see e.g.~\cite{Dorey:2002ik}), this can be seen by
looking at the fermion susy variation
$\delta\lambda\sim F^{\mu \nu}\gamma_{\mu \nu}\epsilon$ (plus terms involving
scalars), where $\epsilon$
is a four-component Dirac spinor and $\gamma_{\mu\nu}$ are proportional to the
generators of the reducible spinorial representation of $SO(4)$ \cite{Blau}.
The latter have the block-diagonal form \eqref{gamma_munu},
where the matrices $\sigma_{\mu \nu}$ and $\bar{\sigma}_{\mu \nu}$ are
anti-selfdual and selfdual respectively, cf.~appendix \ref{conv}. Therefore,
plugging \eqref{instanton} (e.g.~for a selfdual field strength $F^+$) in the
susy variation leads (for vanishing scalars) to
\begin{equation}
\delta\lambda\sim\left(
\begin{array}{cc}
 0 & 0 \\ 0 & (F^+)^{\mu \nu}\bar{\sigma}_{\mu\nu}
\end{array}\right)
\left(\begin{array}{cc}
\epsilon^+ \\ \epsilon^- 
\end{array}\right)\ ,
\end{equation}
showing that the configuration is half-supersymmetric. In the
${\cal N}=1$ theory, the spinors are taken chiral and according to the choice
of chirality either the instanton or the anti-instanton represent maximally supersymmetric
solutions.

Note that in theories with ${\cal N}>1$, many configurations preserving one or
more supersymmetries do not have (anti-)selfdual field strength.
We shall see this explicitely below.

A class of finite-action solutions of euclidean Yang-Mills theory was
explicitly constructed by Belavin et al.~\cite{Belavin:1975fg} (see also
\cite{Vandoren:2008xg,Belitsky:2000ws} for reviews). The field strength is
selfdual and the gauge potential for $k=1$ and gauge group $SU(2)$ (in regular
gauge) takes the form
\begin{equation}\label{eq: A 1 istantone}
A^a_\mu(x;x_0,\rho)=2\frac{\eta^a_{\mu\nu}(x-x_0)^\nu}{(x-x_0)^2+\rho^2}\ ,
\end{equation}
where the arbitrary parameters $x_0^\mu$ and $\rho$ are called collective
coordinates and $\eta^a_{\mu\nu}$ are the 't Hooft symbols defined in appendix
\ref{conv}. Taking into account the gauge orientation, the total number of
collective coordinates in this situation is 8 \cite{Vandoren:2008xg}. One can
show by computing the index of the Dirac operator in an instanton background
(with selfdual field strength) that there are $4Nk$ bosonic collective
coordinates for an instanton with winding number $k$ and gauge group $SU(N)$,
counting the number of solutions to the selfduality equations with fixed
topological charge $k$ \cite{Vandoren:2008xg}.
This calculation also reveals that in the background of an (anti-)instanton,
the Dirac equation can have non-trivial solutions
$\bar\lambda_{cl}$ ($\lambda_{cl}$) only for negative (positive) chirality
spinors, and that the number of such solutions in $2Nk$ \cite{Vandoren:2008xg}.
These zero modes are parametrized by the so-called fermionic collective
coordinates.

In euclidean super Yang-Mills theories, the previous results can
be generalized as follows. First, an obvious solution to the equations of motion
is given by \eqref{instanton} with all the other fields vanishing. Also, when
scalar fields are absent (such as in ${\cal N}=1$ SYM) or uncoupled to
fermions, another solution is given by $F$ (anti-)selfdual,
$\bar\lambda=\bar\lambda_{cl}$ ($\lambda =\lambda_{cl}$) and all remaining
fields vanishing. But as soon as fermion-scalar couplings are turned on, as in
the case of ${\cal N}=4$ SYM, the latter configuration no longer
solves the equations of motion \eqref{eomF}-\eqref{eombarlambda}.

For the sake of definiteness, let us remind an iterative way to construct
solutions when the gauge group is $SU(2)$ \cite{Belitsky:2000ws}. Start from a
configuration $\Phi=(A=A_{cl},\,\phi^{AB}=\lambda^A=\bar\lambda_A=0)$, where
$A_{cl}$ is a gauge potential for a selfdual field strength. A new solution is
obtained by
\begin{equation}\label{eq: superistantone supertrasformazione}
\Phi(\bar\zeta)=e^{i\bar\zeta_A\bar{Q}^A}\Phi=
\sum_{n=0}^\infty\frac1{n!}\delta^n\Phi\ ,
\end{equation}
where the ${\bar Q}^A$ are the susy generators, and the last equality comes from
expanding perturbatively in the fermionic susy parameter $\bar\zeta$. Due to
the anti-selfduality of $\sigma_{\mu\nu}$, the third equation of
\eqref{SusyEqCompl} implies $e^{i\zeta^AQ_A}\Phi=\Phi$, and therefore the
positive chirality susy generators cannot be used to generate from $\Phi$ a
new solution. Using \eqref{SusyEqCompl}, and starting from the configuration
$\Phi$, one successively obtains
\begin{equation}\label{eq: superistantone trasformazione sviluppato}
  \begin{split}
    &\phantom{}^{(0)}\!A=A_{cl}\ , \\
    &\phantom{}^{(1)}\!\bar\lambda_{A}=-\frac12\bar\sigma^{\mu\nu}\bar\zeta_{A}\phantom{}^{(0)}\!F_{\mu\nu}\ , \\
    &\phantom{}^{(2)}\!\phi^{AB}=\frac{1}{\sqrt{2}}\epsilon^{ABCD}\bar\zeta_C\phantom{}^{(1)}\bar\lambda_{D}
        =-\frac1{2\sqrt2}\epsilon^{ABCD}\bar\zeta_C\bar\sigma^{\mu\nu}\bar\zeta_{D}\phantom{}^{(0)}\!F_{\mu\nu}\ , \\
    &\phantom{}^{(3)}\!\lambda^{\alpha,A}=-\frac{i\sqrt2}{3}\bar{\zeta}_{\dot\alpha,B}\sls{D}^{\alpha\dot\alpha}\phantom{}^{(2)}\!\phi^{AB}
        =\frac i6\epsilon^{ABCD}{\sigma_\rho}^{\alpha\dot\beta}\bar{\zeta}_{\dot\beta,B}\left(\bar\zeta_C\bar\sigma^{\mu\nu}\bar\zeta_{D}\right)D^\rho\phantom{}^{(0)}\!F_{\mu\nu}\ , \\
    &\phantom{}^{(4)}\!A_\mu=-\frac i4\bar\zeta_A\bar\sigma_{\mu}\phantom{}^{(3)}\!\lambda^{A}
    =\frac 1{24}\epsilon^{ABCD}\left(\bar\zeta_A\bar\sigma_{\mu\rho}\bar{\zeta}_{B}\right)\left(\bar\zeta_C\bar\sigma^{\sigma\nu}\bar\zeta_{D}\right)D^\rho\phantom{}^{(0)}\!F_{\sigma\nu}\ , \\
    &\ \ \ \ \vdots
  \end{split}
\end{equation}
where the superscript indicates the number of susy parameters $\bar\zeta$
contained in the field. It turns out that $\bar\zeta$ is one of the two
two-component fermionic collective coordinates for the $SU(2)$ gauge group.
The other one is obtained by using superconformal supersymmetry transformation
laws \cite{Belitsky:2000ws}. The fact that all fermionic zero modes can be
generated by means of ordinary supersymmetry and superconformal transformations
is specific to $SU(2)$, and the situation is more involved for $SU(N>2)$
\cite{Belitsky:2000ws}. The solution constructed iteratively in this way is
called \emph{super-instanton}
(cf.~\cite{Belitsky:2000ws,Dorey:2002ik,Bianchi:2007ft} for
reviews).

\section{Spinorial geometry of four-dimensional euclidean space}
\label{spinorial}
In this section we summarize the essential information needed to realize
spinors of Spin$(4)$ in terms of forms \cite{Gillard:2004xq}. For more
details, we refer to \cite{Lawson:1998yr}.
Consider the real vector space $V=\bR^4$ endowed with its canonical scalar
product and orthonormal basis $\{e_1,e_2,e_3,e_4\}$. Define the subspace $U$
spanned by the first two basis elements, $U=\mbox{span}(e_1,e_2)$.
The space of Dirac spinors $\Delta_\mathbb{C}$ is defined as the exterior
algebra of $U\otimes\mathbb{C}$,
\begin{equation}
\Delta_\mathbb{C}=\Lambda^*(U\otimes\mathbb{C})\ ,
\end{equation}
which is nothing else than
$\mbox{Span}_\mathbb{C}(1,e_1,e_2,e_1\wedge e_2 =: e_{12})$. A generic Dirac
spinor is thus written as
\begin{equation}
\lambda=\lambda_0 1+\lambda_1 e_1+\lambda_2
e_2+\lambda_{12}e_{12}\ ,
\end{equation}
and has 4 complex degrees of freedom as it should. The gamma matrices are
represented on $\Delta_\mathbb{C}$ as
\begin{eqnarray}
\gamma_i\lambda&=&e_i\wedge\lambda+e_i\lrcorner\lambda\ , \nonumber \\
\gamma_{i+2}\lambda&=&ie_i\wedge\lambda-ie_i\lrcorner\lambda\ ,
                     \label{gammaaction}
\end{eqnarray}
where $i=1,2$, and the contraction operator $\lrcorner$ is defined though its
action on a k-form as
\begin{equation}
e_i\lrcorner\left(\frac1{k!}\eta_{i_1\ldots i_k}e_{i_1}\wedge\ldots\wedge
e_{i_k}\right)=\frac1{(k-1)!}\eta_{ij_1\ldots j_{k-1}}e_{j_1}\wedge\ldots\wedge
e_{j_{k-1}}\ .
\end{equation}
One easily checks that this representation of the gamma matrices satisfies
the Clifford algebra relations $\{\gamma_{\mu},\gamma_{\nu}\}=2\delta_{\mu\nu}$.
Note that the elements $e_i$ are indistinctly viewed as vectors or forms
according to the objects on which they are acting.
From the definition of the chirality matrix
$\gamma_5=\gamma_1\gamma_2\gamma_3\gamma_4$ and \eqref{gammaaction}, one
readily sees that 
\begin{equation}
  \begin{array}{ll}
    \gamma_51=1\ , & \gamma_5e_{12}=e_{12}\ , \\
    \gamma_5e_1=-e_1\ , & \gamma_5e_2=-e_2\ ,
  \end{array}
\end{equation}
so that the usual split of the space of Dirac spinors into positive and
negative chirality Weyl spinors here amounts to the split
$\Delta_\mathbb{C}=\Delta^+\oplus\Delta^-$ into forms of even degree
$1,e_{12}$ and forms of odd degree $e_1,e_2$. This decomposition is
invariant under the euclidean Lorentz group $SO(4)=SU(2)\times SU(2)$, with
each subspace transforming under a different $SU(2)$ factor.

Let us define the hermitian inner product
\begin{equation}
\langle a^ie_i|b^je_j\rangle=\sum_{i=1}^2{a^*}^ib^i
\end{equation}
on $U\otimes\mathbb{C}$, and then extend it to $\Delta_\mathbb{C}$.
This yields the Spin$(4)$ invariant Dirac inner product on the space of
spinors $\Delta_\mathbb{C}$,
\begin{equation}
D(\eta,\theta) = \langle\eta|\theta\rangle\ .
\end{equation}
It reveals quite useful for practical purposes to switch to another basis for
the gamma matrices, defining
\begin{eqnarray}
&&\Gamma_1=\frac1{\sqrt2}\left(\gamma_1-i\gamma_3\right)\ , \qquad
  \Gamma_{\bar 1}=\frac1{\sqrt2}\left(\gamma_1+i\gamma_3\right)\ , \nonumber \\
&&\Gamma_2=\frac1{\sqrt2}\left(\gamma_2-i\gamma_4\right)\ , \qquad
  \Gamma_{\bar 2}=\frac1{\sqrt2}\left(\gamma_2+i\gamma_4\right)\ .
  \label{eq: base 1 bar1 2 bar2}
\end{eqnarray}
In this new basis, the gamma matrices satisfy
$\{\Gamma_A,\Gamma_B\}=2\eta_{AB}$, $A,B=\{1,\bar{1},2,\bar{2}\}$, where the
non-vanishing components of the metric are $\eta_{1\bar{1}}=\eta_{\bar{1}1}=1$,
$\eta_{2\bar{2}}=\eta_{\bar{2}2}=1$.

The advantage of this new basis stems from the fact that the $\Gamma_A$
satisfy a fermionic annihilation-creation operator algebra, since
$\{\Gamma_1,\Gamma_{\bar{1}}\}=2$, $\{\Gamma_2,\Gamma_{\bar{2}}\}=2$, for which
the spinor $1$ can be identified as the vacuum state, being annihilated by
$\Gamma_{\bar{1}}$ and $\Gamma_{\bar{2}}$:
\begin{equation}\label{CliffordVacuum}
\Gamma_{\bar{1}} 1=\Gamma_{\bar{2}}1 =0\ .
\end{equation}
All the other states can be constructed by acting with $\Gamma_1$ and
$\Gamma_2$ on $1$. Using \eqref{eq: base 1 bar1 2 bar2} and
\eqref{gammaaction}, one
can compute the action of the gamma matrices and the Spin$(4)$ generators
$\Gamma_{AB} = \Gamma_{[A}\Gamma_{B]}$ on the basis spinors. This is summarized
in table \ref{tab:gammabase}, where another simplification coming from the use
of the basis \eqref{eq: base 1 bar1 2 bar2} is apparent from the vanishing of
half of the entries of this table\footnote{Spinors that are annihilated by
half of the gamma matrices are sometimes referred to as \emph{pure spinors}.
Table \ref{tab:gammabase} is a manifestation of the fact that in 6 or less real dimensions,
all spinors are pure.}.

\begin{table}
\begin{center}
\begin{tabular}{|c|cccc|}
  \hline
                            & $1$           & $e_1$             & $e_2$             & $e_{12}$ \\
  \hline
  $\Gamma_1$                & $\sqrt{2}e_1$ & $0$               & $\sqrt{2}e_{12}$  & $0$ \\
  $\Gamma_{\bar{1}}$        & $0$           & $\sqrt{2}$        & $0$               & $\sqrt{2}e_2$ \\
  $\Gamma_2$                & $\sqrt{2}e_2$ & $-\sqrt{2}e_{12}$ & $0$               & $0$ \\
  $\Gamma_{\bar{2}}$        & $0$           & $0$               & $\sqrt{2}$        & $-\sqrt{2}e_1$ \\
  \hline
  \hline
  $\Gamma_{1\bar{1}}$       & $-1$          & $e_1$             & $-e_2$            & $e_{12}$ \\
  $\Gamma_{12}$             & $2e_{12}$     & $0$               & $0$               & $0$ \\
  $\Gamma_{1\bar{2}}$       & $0$           & $0$               & $2e_1$            & $0$ \\
  $\Gamma_{\bar{1}2}$       & $0$           & $-2e_2$           & $0$               & $0$ \\
  $\Gamma_{\bar{1}\bar{2}}$ & $0$           & $0$               & $0$               & $-21$ \\
  $\Gamma_{2\bar{2}}$       & $-1$          & $-e_1$            & $e_2$             & $e_{12}$ \\
  \hline
\end{tabular}
\end{center}
\caption{\small{Action of the gamma matrices and Spin$(4)$ generators on the
basis $1,e_1,e_2,e_{12}$.}}\label{tab:gammabase}
\end{table}

We will sometimes use another basis, in which the gamma matrices are given
by \eqref{rep-gamma}. The spinors $\{1,e_1,e_2,e_{12}\}$ can easily be
expressed in that basis by starting from \eqref{CliffordVacuum} and acting
with the creation operators on the vacuum. Their form is fixed up to a global
phase by normalizing the states to unit norm:
\begin{eqnarray}
&&1^\alpha=\frac1{\sqrt2}\left(
  \begin{array}{c}
    1\\-i
  \end{array}
  \right)\,, \qquad
  e_{12}^\alpha=\frac1{\sqrt2}\left(
  \begin{array}{c}
    i\\-1
  \end{array}
  \right)\,, \nonumber \\
&&{e_1}_{\dot\alpha}=\frac1{\sqrt2}\left(
  \begin{array}{c}
    1\\i
  \end{array}
  \right)\,, \qquad
  {e_2}_{\dot\alpha}=\frac1{\sqrt2}\left(
  \begin{array}{c}
    -i\\-1
  \end{array}
  \right)\,, \label{2-comp}
\end{eqnarray}
where we suppressed two lower zeroes in $1,e_{12}$ and two upper
zeroes in $e_1,e_2$. \eqref{2-comp} provides thus an expression of the
forms in usual two-component notation.
For practical purposes, we also mention their complex conjugates,
\begin{equation}
(1^\alpha)^*=-ie_{12}^{\alpha}\,, \qquad (e_{12}^\alpha)^*=-i1^{\alpha}\,, \qquad
({e_1}_{\dot\alpha})^*=i{e_2}_{\dot\alpha}\,, \qquad ({e_2}_{\dot\alpha})^*=
i{e_1}_{\dot\alpha}\,,
\end{equation}
the corresponding spinors with raised/lowered indices,
\begin{eqnarray}
&&1_\alpha=1^\beta\epsilon_{\beta\alpha}=i1^\alpha\,, \qquad
{e_{12}}_{\alpha}=e_{12}^\beta\epsilon_{\beta\alpha}=-ie_{12}^\alpha\,,
\nonumber \\
&&e_1^{\dot\alpha}=\epsilon^{\dot\alpha\dot\beta}{e_1}_{\dot\beta}=
-i{e_1}_{\dot\alpha}\,, \qquad
e_2^{\dot\alpha}=\epsilon^{\dot\alpha\dot\beta}{e_2}_{\dot\beta}=
i{e_2}_{\dot\alpha}\,, \nonumber
\end{eqnarray}
as well as the various contractions of the basis spinors,
\begin{equation}\label{eq: contrazioni spinoriali}
1^\alpha {e_{12}}_{\alpha}=1\,, \qquad e_{12}^\alpha 1_\alpha=-1\,, \qquad
e_1^{\dot\alpha}{e_2}_{\dot\alpha}=-1\,, \qquad e_2^{\dot\alpha}{e_1}_{\dot\alpha}
=1\,.
\end{equation}

Finally, let us express for further reference the (anti-)selfduality
conditions \eqref{instanton} in terms of the basis
\eqref{eq: base 1 bar1 2 bar2}:
\begin{eqnarray}\label{SDSpinBasis}
F_{\mu\nu} = \frac12\epsilon_{\mu\nu\rho\sigma} F^{\rho \sigma} \quad
\Leftrightarrow \quad F^{1\bar1} + F^{2\bar2} = 0\,, \quad F^{12}=0\,, \\
F_{\mu\nu} = -\frac12\epsilon_{\mu\nu\rho\sigma} F^{\rho \sigma} \quad
\Leftrightarrow \quad F^{1\bar1} - F^{2\bar2} = 0\,, \quad F^{1\bar2}=0\,.
\label{ASDSpinBasis}
\end{eqnarray}
In these coordinates, the complex conjugation simply amounts to change a barred
index into an unbarred one, e.g.~$(V^{\bar i})^*=V^i$, $(F^{i\bar j})^* =
F^{\bar i j}$, etc.
Therefore $F^{1 \bar1}$ and $F^{2 \bar2}$ are purely imaginary, and the
equations \eqref{SDSpinBasis}, \eqref{ASDSpinBasis} each impose three real
conditions on the components of $F^{\mu\nu}$ as they should. \eqref{SDSpinBasis}
and \eqref{ASDSpinBasis} are sometimes referred to as the \emph{Donaldson
equations}. We will encounter in the next section generalizations thereof
including in particular the scalar fields present in ${\cal N}=4$ SYM.

\section{Classification of supersymmetric solutions}\label{Classification}

\subsection{Representatives}\label{Rapp}

The spinorial geometry approach is tailored to fully exploit the linearity of
the Killing spinor equations. One of its basic ingredients is the use of the
symmetries of the theory to transform the Killing spinors to preferred
representatives of their orbit under this symmetry group. This is the scope of
the present subsection.

Using the results of the previous section, the Killing spinors appearing in
\eqref{SusyEqCompl} can be expressed in the language of forms as
\begin{eqnarray}
&&\xi^{\alpha,A}=\omega_0^A1^{\alpha}+\omega_{12}^Ae_{12}^{\alpha}\,, \nonumber \\
&&\bar\xi_{\dot\alpha,A}=\omega_{1,A}{e_1}_{\dot\alpha}+\omega_{2,A}
{e_2}_{\dot\alpha}\,, \label{Generic Killing Spinor}
\end{eqnarray}
where the coefficients $\omega^A$ ($A=1,\ldots,4$) are complex numbers that
are related by the symplectic Majorana condition \eqref{eq: simplettica},
which imposes the following structure:
\begin{equation}
\omega_0=\left(
    \begin{array}{c}
      a\\b\\c\\d
    \end{array}
    \right)\,, \qquad
    \omega_{12}=\left(
    \begin{array}{c}
      -d^*\\-c^*\\b^*\\a^*
    \end{array}
    \right)\,, \qquad
\omega_1=\left(
    \begin{array}{c}
      e\\f\\g\\h
    \end{array}
    \right)\,, \qquad
    \omega_2=\left(
    \begin{array}{c}
      h^*\\g^*\\-f^*\\-e^*
    \end{array}
    \right)\,. \label{eq: struttura simplettici}
\end{equation}
This structure is preserved both by the $SO(4)$ Lorentz and by the $SO(5,1)$
internal R-symmetry transformations \cite{Vandoren:2008xg}.
We are now going to use the latter symmetries to simplify the form
\eqref{Generic Killing Spinor}-\eqref{eq: struttura simplettici} of a generic
Killing spinor. These are generated by the $\gamma_{\mu\nu}$ and
${\hat\gamma}_{ab}$ given respectively in \eqref{gamma_munu} and
\eqref{hatgamma_ab}.

First, we notice that for a right chiral spinor there exists a single orbit
for the action of $SO(5,1)$, and consequently a single representative that can
be brought to the form
\begin{equation}\label{eq: rappresentativo chirale}
  \xi_1=
  \left(
  \begin{array}{c}
    1\\0\\0\\0
  \end{array}
  \right)1+
  \left(
  \begin{array}{c}
    0\\0\\0\\1
  \end{array}
  \right)e_{12}\,.
\end{equation}
To see this, it is enough to find a series of transformations bringing the
vector $(1,0,0,0)^T$ to $(a,b,c,d)^T$, for arbitrary $a,b,c,d \in \mathbb{C}$.
This is done for instance by
\begin{equation}
  e^{\alpha\Sigma^{25}}e^{\beta\Sigma^{14}}
  \left(
  \begin{array}{c}
    1\\0\\0\\0
  \end{array}
  \right)=
  \left(
  \begin{array}{c}
    e^{-i\alpha} e^{\beta}\\0\\0\\0
  \end{array}
  \right)=
  \left(
  \begin{array}{c}
    a_1\\0\\0\\0
  \end{array}
  \right)\,,
\end{equation}
with $a_1 \in \mathbb{C}$, followed by
\begin{equation}
  e^{\alpha\Sigma^{25}}e^{\beta\Sigma^{45}}
  \left(
  \begin{array}{c}
    a_1\\0\\0\\0
  \end{array}
  \right)=
  \left(
  \begin{array}{c}
    a_1\cos{\beta}e^{-i\alpha}\\-a_1\sin{\beta}e^{i\alpha}\\0\\0
  \end{array}
  \right)=
  \left(
  \begin{array}{c}
    a_2\\b_2\\0\\0
  \end{array}
  \right)\,,
\end{equation}
where $a_2,b_2\in\mathbb{C}$ are independent, and finally
\begin{displaymath}
e^{\frac{\alpha}2[\Sigma^{25}-\Sigma^{36}]}e^{\frac{\beta}2[\Sigma^{23}+\Sigma^{56}]}
e^{\frac{\gamma}2[\Sigma^{25}+\Sigma^{36}]}e^{\frac{\delta}2[\Sigma^{23}-\Sigma^{56}]}
  \left(
  \begin{array}{c}
    a_2\\b_2\\0\\0
  \end{array}
  \right)=
  \left(
  \begin{array}{c}
    a_2\cos{\delta}e^{-i\gamma}\\b_2\cos{\beta}e^{i\alpha}\\-b_2\sin{\beta}
    e^{-i\alpha}\\-a_2\sin{\delta}e^{i\gamma}
  \end{array}
  \right)=
  \left(
  \begin{array}{c}
    a\\b\\c\\d
  \end{array}
  \right)\,.
\end{displaymath}
Similarly, there exists a single orbit, and hence a single representative
under $SO(5,1)$ for left chiral spinors.

After having brought a generic chiral Killing spinor to a simpler form
(\eqref{eq: rappresentativo chirale} for a right spinor), one looks for the
subgroup of the global symmetry group that leaves it invariant. This stability
subgroup can then be used to simplify additional Killing spinors. For the
representative \eqref{eq: rappresentativo chirale}, one checks that the
$SO(5,1)$ generators stabilizing it are
\begin{equation}
S_1=\frac12(\hat\gamma^{25}-\hat\gamma^{36})\,, \qquad
S_2=\frac12(\hat\gamma^{23}+\hat\gamma^{56})\,, \qquad
S_3=\frac12(\hat\gamma^{26}+\hat\gamma^{35})\,,
\label{eq: sottogruppo stabilita SO(5,1)}
\end{equation}
\begin{displaymath}
S_4=\frac12(\hat\gamma^{15}+\hat\gamma^{45})\,, \qquad
S_5=\frac12(\hat\gamma^{12}-\hat\gamma^{24})\,, \qquad
S_6=\frac12(\hat\gamma^{16}+\hat\gamma^{46})\,, \qquad
S_7=\frac12(\hat\gamma^{13}-\hat\gamma^{34})\,,
\end{displaymath}
where $S_1$, $S_2$ and $S_3$ form an $su(2)$ subalgebra, while $S_4,\cdots,S_7$
form an abelian ideal. Being a right chiral spinor of Spin$(4)$, it is also
stabilized by the $SU(2)_L$ subgroup of Spin$(4)$ generated by
$\Gamma_{1\bar1}-\Gamma_{2\bar2}$, $\Gamma_{1\bar2}$ and
$\Gamma_{\bar12}$. On the other hand, the $SU(2)_R$ generators
$\Gamma_{1\bar1}+\Gamma_{2\bar2}$, $\Gamma_{12}$ and $\Gamma_{\bar1\bar2}$
only mix the forms $1$ and $e_{12}$ (Spin$(4)$ transformations preserve
chirality), so that applying an $SU(2)_R$ transformation to the representative
\eqref{eq: rappresentativo chirale} yields again a spinor of the form
$\omega_01+\omega_{12}e_{12}$, with $\omega_0$, $\omega_{12}$ given in
\eqref{eq: struttura simplettici}. But, as we just explained, that spinor is
in the same orbit as \eqref{eq: rappresentativo chirale} under $SO(5,1)$.
The action of the $SU(2)_R$ can thus be compensated by a subsequent
$SO(5,1)$ transformation. It is easy to see that the $SO(5,1)$ generators
accomplishing this are
\begin{equation}
S_8 = \frac12(\hat\gamma^{25}+\hat\gamma^{36})\,, \qquad
S_9 = \frac12(\hat\gamma^{23}-\hat\gamma^{56})\,, \qquad
S_{10} = \frac12(\hat\gamma^{26}-\hat\gamma^{35})\,,
\end{equation}
which of course span an $su(2)$ algebra that commutes with the one
formed by $S_1,\ldots,S_3$. Together, $S_1,\ldots,S_{10}$ generate the Euclidean
group $SO(4)\ltimes\bR^4\cong ISO(4)$, so that the stability subgroup of
\eqref{eq: rappresentativo chirale} is $SU(2)_L\times ISO(4)$.

Let us see how a second Killing spinor can be simplified using the stability
subgroup of the first. It can have either the same chirality, or the opposite
one\footnote{Here, we will always consider Killing spinors with a definite
chirality. This is restrictive, but as far as instantons are concerned, it is
the most important case.}. First assume that the second one has the same
chirality. It can thus be written as
\begin{equation}
  \xi_2=\left(
  \begin{array}{c}
    a\\b\\c\\d
  \end{array}
  \right)1+
  \left(
  \begin{array}{c}
    -d^*\\-c^*\\b^*\\a^*
  \end{array}
  \right)e_{12}\,.
\end{equation}
One observes that when the second spinor is of the form
\begin{equation}\label{(2000)}
  \xi_2^{(1)}=\left(
  \begin{array}{c}
    a\\0\\0\\d
  \end{array}
  \right)1+
  \left(
  \begin{array}{c}
    -d^*\\0\\0\\a^*
  \end{array}
  \right)e_{12}\,,
\end{equation}
we may use an $SU(2)_R$ transformation (compensated by $S_8,\ldots,S_{10}$)
to choose $a,d\in\bR$. The residual isotropy group leaving invariant
\eqref{eq: rappresentativo chirale} as well as \eqref{(2000)} is then given by
$SU(2)_L\times((U(1)\times SU(2))\ltimes\bR^4)$, with the $U(1)$ generated
by $S_9$. \eqref{(2000)} actually belongs to the same Lorentz orbit
as \eqref{eq: rappresentativo chirale}, in that an arbitrary right $SO(4)$
chiral spinor $a1 - d^* e_{12}$ can always be brought to the spinor $\rho^21$,
$\rho\in \mathbb{R}$ by means of an $SO(4)$ transformation (to show this,
consider e.g. $\mbox{exp}(\alpha\gamma^{13})\mbox{exp}(\beta\gamma^{23})
\mbox{exp}(\delta\gamma^{13})$ acting on $\rho^21$).
Actually, since the Killing spinor equations \eqref{SusyEqCompl}
are linear in the spinorial parameter, we will generally forget about the
scaling factor $\rho^2$.

On the other hand, when $b$ or $c$ are different from zero, the second spinor
can, up to an overall factor, be brought to the form
\begin{equation}\label{(1100)}
  \xi_2^{(2)}=\left(
  \begin{array}{c}
    0\\1\\0\\0
  \end{array}
  \right)1+
  \left(
  \begin{array}{c}
    0\\0\\1\\0
  \end{array}
  \right)e_{12}\,,
\end{equation}
using the generators $S_1,\ldots,S_7$ (this is obvious by acting on
\eqref{(1100)} with $e^{\alpha S_1}e^{\beta S_2}e^{\gamma S_1}e^{\delta S_7}$
$e^{\eta S_6}e^{\theta S_5}e^{\Omega S_4}$). The spinor (\ref{(1100)}) is further stabilized by $S_8$, $S_9$ and $S_{10}$.

We will use the following pictorial representation. Without loss of generality,
the first chiral Killing spinor can always be taken of the form
\eqref{eq: rappresentativo chirale}. Next, a configuration admitting as Killing
spinors \eqref{eq: rappresentativo chirale} and \eqref{(2000)} will be
represented by
\begin{center}
\scalebox{.9}{\begin{picture}(30,30)(30,0)
\multiframe(0,10)(10.5,0){1}(10,10){\footnotesize$\bullet$}
\multiframe(10,10)(10.5,0){1}(10,10){\footnotesize$\bullet$}
\multiframe(20,10)(10.5,0){1}(10,10){}
\multiframe(30,10)(10.5,0){1}(10,10){}

\multiframe(0,-0.5)(10.5,0){1}(10,10){}
\multiframe(10,-0.5)(10.5,0){1}(10,10){}
\multiframe(20,-0.5)(10.5,0){1}(10,10){}
\multiframe(30,-0.5)(10.5,0){1}(10,10){}

\end{picture}}\hspace*{-0.5cm}, 
\end{center}
while a configuration admitting as Killing spinors
\eqref{eq: rappresentativo chirale} and \eqref{(1100)} shall be denoted by
\begin{center}
\scalebox{.9}{\begin{picture}(30,30)(30,0)
\multiframe(0,10)(10.5,0){1}(10,10){\footnotesize$\bullet$}
\multiframe(10,10)(10.5,0){1}(10,10){}
\multiframe(20,10)(10.5,0){1}(10,10){}
\multiframe(30,10)(10.5,0){1}(10,10){}

\multiframe(0,-0.5)(10.5,0){1}(10,10){\footnotesize$\bullet$}
\multiframe(10,-0.5)(10.5,0){1}(10,10){}
\multiframe(20,-0.5)(10.5,0){1}(10,10){}
\multiframe(30,-0.5)(10.5,0){1}(10,10){}

\end{picture}}\hspace*{-0.5cm}.
\end{center}
 
Each dot represents a Killing spinor. A solution can have at most 8 real
Killing spinors of the same chirality, here right. Spinors that can be related
by a Spin$(4)$ transformation are written on the same line. In the
second case, for example, the two Killing spinors can only be related by an
internal $SO(5,1)$ transformation.

One can also consider Killing spinors of different chiralities. Let us again
fix the first spinor to be of the form \eqref{eq: rappresentativo chirale},
and use its stability subgroup to simplify a left chiral spinor
\begin{equation}\label{LeftChiralSpinor}
  \bar\xi=
  \left(
  \begin{array}{c}
    e\\f\\g\\h
  \end{array}
  \right)e_1+
  \left(
  \begin{array}{c}
    h^*\\g^*\\-f^*\\-e^*
  \end{array}
  \right)e_2\,.
\end{equation}
There are again two cases. If $e$ or $h$ is different from zero, one can use
the Lorentz subgroup $SU(2)_L$ as well as $S_4,\ldots,S_7$ to bring the
left chiral spinor to
\begin{equation}\label{LeftChiral1}
  \bar\xi_2^{(1)}=
  \left(
  \begin{array}{c}
    \rho^2\\0\\0\\0
  \end{array}
  \right)e_1+
  \left(
  \begin{array}{c}
    0\\0\\0\\-\rho^2
  \end{array}
  \right)e_2\,.
\end{equation}
This spinor is still invariant under $SO(4)$ transformations generated by
$S_1,\ldots,S_3$ and $S_8,\ldots,S_{10}$.
On the other hand, if $e=h=0$, the generators of
the $SU(2)_L$ subgroup (or equivalently the generators $S_1$, $S_2$ and $S_3$)
can be used to cast it into the form
\begin{equation}\label{LeftChiral2}
  \bar\xi_2^{(2)}=
  \left(
  \begin{array}{c}
    0\\\rho^2\\0\\0
  \end{array}
  \right)e_1+
  \left(
  \begin{array}{c}
    0\\0\\-\rho^2\\0
  \end{array}
  \right)e_2\,,
\end{equation}
which is still stabilized by $S_4,\ldots,S_{10}$. Note that \eqref{LeftChiral1}
and \eqref{LeftChiral2} are actually related by $SO(5,1)$ transformations, but
not of the type \eqref{eq: sottogruppo stabilita SO(5,1)} stabilizing
\eqref{eq: rappresentativo chirale}. This actually shows that a generic Dirac
spinor can be brought either to the form
\eqref{eq: rappresentativo chirale}+\eqref{LeftChiral1} or to
\eqref{eq: rappresentativo chirale}+\eqref{LeftChiral2}, i.e.~there are two
orbits under the Lorentz and internal symmetry groups for a generic Killing
spinor.

Allowing for Killing spinors of different chiralities, a solution can preserve
at most 16 real supersymmetries. Again, we will make use of a pictorial
representation to visualize the different possible supersymmetric
configurations in an easy way. A solution admitting
\eqref{eq: rappresentativo chirale} and \eqref{LeftChiral1} as Killings spinors
will be written as
\begin{table}[h]
\begin{center}
\scalebox{.9}{\begin{picture}(30,30)(30,0)
\multiframe(0,10)(10.5,0){1}(10,10){\footnotesize$\bullet$}
\multiframe(10,10)(10.5,0){1}(10,10){}
\multiframe(20,10)(10.5,0){1}(10,10){}
\multiframe(30,10)(10.5,0){1}(10,10){}

\multiframe(0,-0.5)(10.5,0){1}(10,10){}
\multiframe(10,-0.5)(10.5,0){1}(10,10){}
\multiframe(20,-0.5)(10.5,0){1}(10,10){}
\multiframe(30,-0.5)(10.5,0){1}(10,10){}

\multiframe(0,-11)(10.5,0){1}(10,10){\footnotesize$\bullet$}
\multiframe(10,-11)(10.5,0){1}(10,10){}
\multiframe(20,-11)(10.5,0){1}(10,10){}
\multiframe(30,-11)(10.5,0){1}(10,10){}

\multiframe(0,-21.5)(10.5,0){1}(10,10){}
\multiframe(10,-21.5)(10.5,0){1}(10,10){}
\multiframe(20,-21.5)(10.5,0){1}(10,10){}
\multiframe(30,-21.5)(10.5,0){1}(10,10){}

\end{picture} $\equiv (1,0,1,0)$\,,}
\end{center}
\end{table}
\newline
while one admitting \eqref{eq: rappresentativo chirale} and
\eqref{LeftChiral2} is denoted by
\begin{table}[h]
\begin{center}
\scalebox{.9}{\begin{picture}(30,30)(30,0)
\multiframe(0,10)(10.5,0){1}(10,10){\footnotesize$\bullet$}
\multiframe(10,10)(10.5,0){1}(10,10){}
\multiframe(20,10)(10.5,0){1}(10,10){}
\multiframe(30,10)(10.5,0){1}(10,10){}

\multiframe(0,-0.5)(10.5,0){1}(10,10){}
\multiframe(10,-0.5)(10.5,0){1}(10,10){}
\multiframe(20,-0.5)(10.5,0){1}(10,10){}
\multiframe(30,-0.5)(10.5,0){1}(10,10){}

\multiframe(0,-11)(10.5,0){1}(10,10){}
\multiframe(10,-11)(10.5,0){1}(10,10){}
\multiframe(20,-11)(10.5,0){1}(10,10){}
\multiframe(30,-11)(10.5,0){1}(10,10){}

\multiframe(0,-21.5)(10.5,0){1}(10,10){\footnotesize$\bullet$}
\multiframe(10,-21.5)(10.5,0){1}(10,10){}
\multiframe(20,-21.5)(10.5,0){1}(10,10){}
\multiframe(30,-21.5)(10.5,0){1}(10,10){}

\end{picture} $\equiv (1,0,0,1)$\,.}
\end{center}
\end{table}

We make use of the shorthand notation $(n_1,n_2,n_3,n_4)$ to indicate the
number $n_i$ of Killing spinors lying in the $i^{\text{th}}$ Spin$(4)$ orbit,
whose representatives are given respectively by
\eqref{eq: rappresentativo chirale},
\eqref{(1100)}, \eqref{LeftChiral1}, \eqref{LeftChiral2}. There are obviously
symmetries of these diagrams that would yield equivalent configurations. For
example, up to the choice of chirality of the first Killing spinor, the
configurations obtained by exchanging the first two lines with the last two
ones are equivalent, e.g. $(2,0,1,0) = (1,0,2,0)$, $(4,0,0,0) = (0,0,4,0)$ etc.

\subsection{Bosonic configurations}\label{Bosons}
Let us now use the machinery of the previous section to classify the purely
bosonic configurations. We therefore put all fermions to zero
in \eqref{SusyEqCompl}, and are thus left with
\begin{eqnarray}
&&\delta\lambda^{\alpha,A}=-\frac12\sigma^{\mu\nu\alpha}_{\ \ \ \ \beta}F_{\mu\nu}
\xi^{\beta,A}-i\sqrt2\bar{\xi}_{\dot\alpha,B}\sls{D}^{\alpha\dot\alpha}\phi^{AB}+
[\phi^{AB},\bar\phi_{BC}]\xi^{\alpha,C}=0\,, \nonumber \\
&&\delta\bar\lambda_{\dot\alpha,A}=-\frac12\bar\sigma^{\mu\nu\ \dot\beta}_{\ \
\dot\alpha}F_{\mu\nu}\bar\xi_{\dot\beta,A}+i\sqrt2\xi^{\alpha,B}
\sls{D}_{\dot\alpha\alpha}\bar\phi_{AB}+[\bar\phi_{AB},\phi^{BC}]
\bar\xi_{\dot\alpha,C}=0\,. \label{eq: invarianza configurazione bosonica}
\end{eqnarray}
These equations can be rewritten to emphasize the action of the operators in
table \ref{tab:gammabase} and avoiding explicit spinor indices:
\begin{eqnarray}
-\frac12F^{\mu\nu}\Gamma_{\mu\nu}\xi^{A}+\sqrt2\Gamma_\mu\bar\xi_BD^\mu\phi^{AB}+
[\phi^{AB},\bar\phi_{BC}]\xi^C&=&0\,, \nonumber \\
-\frac12F^{\mu\nu}\Gamma_{\mu\nu}\bar\xi_{A}+\sqrt2\Gamma_\mu\xi^BD^\mu
\bar\phi_{AB}+[\bar\phi_{AB},\phi^{BC}]\bar\xi_{C}&=&0\,.
\label{eq: invarianza configurazione bosonica spinoriale}
\end{eqnarray}
Plugging \eqref{Generic Killing Spinor} into these equations, using table
\ref{tab:gammabase}, and requiring the coefficients of $1$, $e_1$, $e_2$ and
$e_{12}$ to vanish yields the following system of 16 complex equations:
\begin{eqnarray}
\omega_0^A(F^{1\bar{1}}+F^{2\bar{2}})+2\omega_{12}^AF^{\bar{1}\bar{2}}+2D^{\bar{1}}
\phi^{AB}\omega_{1,B}+2D^{\bar{2}}\phi^{AB}\omega_{2,B}+[\phi^{AB},\bar\phi_{BC}]
\omega_0^C&=&0\,, {\hspace*{1cm}}\nonumber \\
-\omega_{12}^A(F^{1\bar{1}}+F^{2\bar{2}})-2\omega_{0}^AF^{12}+2D^1\phi^{AB}
\omega_{2,B}-2D^2\phi^{AB}\omega_{1,B}+[\phi^{AB},\bar\phi_{BC}]\omega_{12}^C
&=&0\,, \nonumber \\
-\omega_{1,A}(F^{1\bar{1}}-F^{2\bar{2}})-2\omega_{2,A}F^{1\bar{2}}+2D^1
\bar\phi_{AB}\omega_0^B-2D^{\bar{2}}\bar\phi_{AB}\omega_{12}^B+[\bar\phi_{AB},
\phi^{BC}]\omega_{1,C}&=&0\,, \nonumber \\
\omega_{2,A}(F^{1\bar{1}}-F^{2\bar{2}})+2\omega_{1,A}F^{\bar{1}2}+2D^{\bar{1}}
\bar\phi_{AB}\omega_{12}^B+2D^2\bar\phi_{AB}\omega_0^B+[\bar\phi_{AB},
\phi^{BC}]\omega_{2,C}&=&0\,. \label{eq: vincoli geometria spinoriale}
\end{eqnarray}
However, due to the conditions \eqref{eq: struttura simplettici},
the equations in the first and third line of
\eqref{eq: vincoli geometria spinoriale} are related to those in the
second and fourth line by complex conjugation.
We are thus left with 16 independent real equations.

Without loss of generality, one can choose the first Killing spinor to be
\eqref{eq: rappresentativo chirale}, for which the system
\eqref{eq: vincoli geometria spinoriale} boils down to
\begin{eqnarray}
&&F^{1\bar{1}}+F^{2\bar{2}}-[\phi^{12},\phi^{34}]+[\phi^{13},\phi^{24}]=0\,,
\nonumber \\
&&F^{\bar{1}\bar{2}}+[\phi^{24},\phi^{34}]=0\,, \nonumber \\
&&D^{1}\phi^{34}-D^{\bar{2}}\phi^{13}=0\,, \nonumber \\
&&D^{1}\phi^{24}-D^{\bar{2}}\phi^{12}=0\,, \label{Eq.FirstSpinor}
\end{eqnarray}
as well as
\begin{equation}
[\phi^{23},\phi^{AB}]=D^{\mu}\phi^{23}=0\,. \label{phi23decoupl}
\end{equation}
If we define a Lie-algebra valued one-form $\phi$ (the "Higgs field")
with components
\begin{equation}
\phi_1 = (\phi_{\bar 1})^* = \phi^{34}\ , \qquad \phi_2 = (\phi_{\bar 2})^* =
\phi^{24}\,,
\end{equation}
the system \eqref{Eq.FirstSpinor} can be rewritten in the simple form
\begin{equation}
(F-\phi\wedge\phi)^{1\bar 1} + (F-\phi\wedge\phi)^{2\bar 2} = 0\,, \qquad
(F-\phi\wedge\phi)^{12} = 0\,, \label{gen-selfdual}
\end{equation}
\begin{equation}
(D\phi)_{1\bar 1} - (D\phi)_{2\bar 2} = 0\,, \qquad (D\phi)_{1\bar 2} = 0\,,
\label{Dphi-antiselfdual}
\end{equation}
together with
\begin{equation}
D\star\phi = 0\,, \label{Dstarphi}
\end{equation}
where $\star\phi$ denotes the Hodge dual of $\phi$. \eqref{gen-selfdual}
means that the combination $F-\phi\wedge\phi$ must be selfdual, while
\eqref{Dphi-antiselfdual} is nothing else than the
anti-selfduality condition for $D\phi$. Similar equations appeared previously
in \cite{Bonelli:1999it,Kapustin:2006pk}\footnote{Ref.~\cite{Kapustin:2006pk}
deals with a twisted version of ${\cal N}=4$ SYM that is relevant to the
geometric Langlands program. This gives a family of topological field theories
parametrized by some $t$ that takes values in the one-dimensional complex
projective space. Taking $t\to\infty$ in eqns.~(3.29)
of \cite{Kapustin:2006pk} yields our system
\eqref{gen-selfdual}-\eqref{Dstarphi}.}.
The reason of why four of
the scalars combine to a one-form (which at first sight appears to transform
differently under Lorentz transformations) lies in the stability subgroup
of the spinor \eqref{eq: rappresentativo chirale}: As was explained in
section \ref{Rapp}, once we fix the representative
\eqref{eq: rappresentativo chirale}, we are free to do
$SO(4)\cong SU(2)_L\times SU(2)_R$ Lorentz rotations only
if the $SU(2)_R$ is compensated by a subsequent $SO(5,1)$ transformation,
and the scalar fields do transform under the latter. A similar situation
occurs in twisted theories.

Using the complex-valued connection
\begin{equation}
{\cal A} = A + i\phi\,,
\end{equation}
\eqref{gen-selfdual} and \eqref{Dphi-antiselfdual} are equivalent to
\begin{equation}
{\cal F} = \star\bar{\cal F}\,, \label{selfduallike}
\end{equation}
where $\cal F$ is the field strength of $\cal A$ and $\bar{\cal F}$ denotes
its complex conjugate. \eqref{gen-selfdual}-\eqref{Dstarphi} bear some
resemblance to the Hitchin equations \cite{Hitchin:1986vp}
\begin{eqnarray}
&&F - \phi\wedge\phi = 0\,, \nonumber \\
&&D\phi = D\star\phi = 0\,. \label{hitchin}
\end{eqnarray}
Note however that in \eqref{hitchin}, $A$ is a connection on a
$G$-bundle $E\to C$, with $C$ a Riemann surface, while in our context,
$A$ is a connection on a bundle over four-dimensional euclidean space.
Moreover, \eqref{hitchin} imply that ${\cal A}=A+i\phi$ is flat,
whereas here $\cal F$ satisfies the selfduality-like condition
\eqref{selfduallike}. Notice also that the Hitchin equations arise by
reduction of the selfduality equations from four to two
dimensions \cite{Hitchin:1986vp}\footnote{Reduction of the
selfduality equations from four to three dimensions yields the monopole
equations. In gravity, the four-dimensional self-duality equations with one
Killing direction imply the 3d Einstein-Weyl equations \cite{hitch-grav}.}.
This raises the question of whether the system
\eqref{gen-selfdual}-\eqref{Dstarphi} also has a higher-dimensional origin.
This is indeed the case: Consider the higher-dimensional analogue of the
selfduality equations \cite{Corrigan:1982th},
\begin{equation}
\frac12 T_{\mu\nu\rho\sigma}F^{\rho\sigma} = \lambda F_{\mu\nu}\,,
\end{equation}
where $\lambda$ is a number and the tensor $T_{\mu\nu\rho\sigma}$ is totally
antisymmetric. If the dimension $D$ is higher than four, $T_{\mu\nu\rho\sigma}$
cannot be invariant under $SO(D)$ anymore. The authors of
\cite{Corrigan:1982th} classified all possible choices for
$T_{\mu\nu\rho\sigma}$ up to $D=8$, requiring that $T_{\mu\nu\rho\sigma}$ be
invariant under a maximal subgroup of $SO(D)$. They found that the case
$D=8$ is of particular interest, because it generalizes most closely the
concept of four-dimensional selfduality. For $D=8$ and the choice
Spin$(7)$ as maximal subgroup of $SO(8)$ there are two possible eigenvalues
$\lambda=1$ and $\lambda=-3$ \cite{Corrigan:1982th}. The former leads to the
set of seven equations\footnote{With respect to \cite{Corrigan:1982th}, we
interchanged the 1- and 3-directions.}
\begin{eqnarray}
F_{32}+F_{14}+F_{56}+F_{78}&=&0\,, \nonumber \\
F_{31}+F_{42}+F_{57}+F_{86}&=&0\,, \nonumber \\
F_{34}+F_{21}+F_{76}+F_{85}&=&0\,, \nonumber \\
F_{35}+F_{62}+F_{71}+F_{48}&=&0\,, \nonumber \\
F_{36}+F_{25}+F_{18}+F_{47}&=&0\,, \nonumber \\
F_{37}+F_{82}+F_{15}+F_{64}&=&0\,, \nonumber \\
F_{38}+F_{27}+F_{61}+F_{54}&=&0\,, \label{octonionic}
\end{eqnarray}
called octonionic instanton equations, since they can be rephrased using
the structure constants of the octonions \cite{Corrigan:1982th}.
Now decompose the vector potential as
\begin{displaymath}
A_M=(A_{\mu},-\phi_3,\phi_2,\phi_1,-\phi_4)\,,
\end{displaymath}
where $M=1,\ldots,8$ and
$\mu=1,\ldots,4$, and suppose that the fields $A_{\mu}$, $\phi_{\mu}$ are
independent of the coordinates $x^5,\ldots,x^8$. Then the octonionic
instanton equations \eqref{octonionic} yield exactly the
system \eqref{gen-selfdual}-\eqref{Dstarphi}\footnote{The relation of
BPS equations in euclidean ${\cal N}=4$ SYM to the octonionic instanton
equations was noticed before in \cite{Bonelli:1999it} for the case of
two active scalars.}.

Looking at \eqref{phi23decoupl}, we observe that the field
$\phi^{23}$ is covariantly constant and commutes with any other scalar.
In what follows, we will refer to these conditions as \emph{decoupling
conditions}. Moreover, the field $\phi^{14}$ does not appear in the susy
equations (except through the fact that it commutes with $\phi^{23}$).
This implies that the ghost $i(\phi^{14}+\phi^{23})$ decouples from the
other fields in the supersymmetry constraints.

Comparing \eqref{Eq.FirstSpinor} to \eqref{SDSpinBasis}, we see
that in presence of scalars the field strength $F$ is no
longer selfdual, but the complex field strength $\cal F$ does obey the
selfduality-like equation \eqref{selfduallike}.
It would be interesting to see if one can use \eqref{selfduallike} to construct
generalizations of instantons to include nonvanishing
scalars\footnote{A particular type of such solutions, termed ic-instantons,
was constructed explicitely in \cite{Bonelli:1999it}.}.

We shall now analyze what happens when requiring the existence of more Killing
spinors.

\subsubsection{Killing spinors of same chirality}\label{same}
One can add a further Killing spinor of the same chirality to the original
configuration $(1,0,0,0)$ in two different ways, either one in the same Lorentz
orbit as the first, namely $(2,0,0,0)$, or the other one $(1,1,0,0)$.

Let us first consider the former case, i.e., we take a second spinor of
the form \eqref{(2000)}. Equations
\eqref{Eq.FirstSpinor} are supplemented with
\begin{eqnarray}
&&\mbox{Im}(d[\phi^{12},\phi^{13}])=0\,, \nonumber \\
&&d([\phi^{12},\phi^{34}]-[\phi^{13},\phi^{24}])+2i\mbox{Im}(a)
[\phi^{24},\phi^{34}]=0\,, \nonumber \\
&&-2i\mbox{Im}(a)D^{\bar{2}}\phi^{13}-d^*D^{\bar{2}}\phi^{34}-dD^{1}\phi^{13}
=0\,, \nonumber \\
&&2i\mbox{Im}(a)D^{\bar{2}}\phi^{12}+d^*D^{\bar{2}}\phi^{24}+dD^{1}\phi^{12}
=0\,. \label{eq: vincoli secondo killing primo tipo}
\end{eqnarray}
From these equations, one easily gets the constraints coming from the presence
of more Killing spinors on the same Lorentz orbit, i.e.~configurations
$(3,0,0,0)$ and $(4,0,0,0)$. First, one may verify that the constraints coming
by considering an additional third Killing spinor forces the field strength to
be selfdual, while giving further constraints on the
covariant derivatives and commutators of the scalar fields. For $(4,0,0,0)$,
one may combine \eqref{Eq.FirstSpinor} with the equations obtained from
\eqref{eq: vincoli secondo killing primo tipo} with $(a,d)=(i,0),(0,1),(0,i)$
which generate a basis for the entire orbit. We obtain the following conditions:
\begin{equation}\label{4000} (4,0,0,0) \qquad \Leftrightarrow \qquad
\left\{\begin{array}{lll}
 F^{1\bar1} + F^{2\bar2} = 0\,, \qquad F^{12}=0\,,  \\[.3em]
  [\phi^{23} , \phi^{AB}] = 0  \qquad \forall \; A,B\,, \\[.3em]
  D^\mu \phi^{AB} = 0 \qquad\forall\; A,B\;\mbox{but for}\; \phi^{14}\,, \\[.3em]
 [\phi^{24},\phi^{34}] = [\phi^{12},\phi^{13}] =0\,, \\[.3em]
 [\phi^{12},\phi^{34}] = [\phi^{13},\phi^{24}]\,.\end{array}\right .
\end{equation}
The scalar field $\phi^{14}$ is thus the only field which is left completely
unconstrained (except from its vanishing commutator with $\phi^{23}$).
Adding one more chiral Killing spinor to $(4,0,0,0)$ directly leads to the
instanton solution with selfdual field strength, and vanishing covariant
derivatives and commutators for all scalar fields. It is clear from
\eqref{eq: invarianza configurazione bosonica} that this solution preserves
8 real supersymmetries, thus
\begin{center}
\scalebox{.9}{\begin{picture}(30,30)(30,0)
\multiframe(0,10)(10.5,0){1}(10,10){\footnotesize$\bullet$}
\multiframe(10,10)(10.5,0){1}(10,10){\footnotesize$\bullet$}
\multiframe(20,10)(10.5,0){1}(10,10){\footnotesize$\bullet$}
\multiframe(30,10)(10.5,0){1}(10,10){\footnotesize$\bullet$}

\multiframe(0,-0.5)(10.5,0){1}(10,10){\footnotesize$\bullet$}
\multiframe(10,-0.5)(10.5,0){1}(10,10){}
\multiframe(20,-0.5)(10.5,0){1}(10,10){}
\multiframe(30,-0.5)(10.5,0){1}(10,10){}
\end{picture}}
$\Rightarrow$
\scalebox{.9}{\begin{picture}(30,30)(-20,0)
\multiframe(0,10)(10.5,0){1}(10,10){\footnotesize$\bullet$}
\multiframe(10,10)(10.5,0){1}(10,10){\footnotesize$\bullet$}
\multiframe(20,10)(10.5,0){1}(10,10){\footnotesize$\bullet$}
\multiframe(30,10)(10.5,0){1}(10,10){\footnotesize$\bullet$}

\multiframe(0,-0.5)(10.5,0){1}(10,10){\footnotesize$\bullet$}
\multiframe(10,-0.5)(10.5,0){1}(10,10){\footnotesize$\bullet$}
\multiframe(20,-0.5)(10.5,0){1}(10,10){\footnotesize$\bullet$}
\multiframe(30,-0.5)(10.5,0){1}(10,10){\footnotesize$\bullet$}
\end{picture}}
\end{center}

We now focus on the latter case, i.e.~$(1,1,0,0)$. Combining
\eqref{Eq.FirstSpinor} with the
constraints coming from \eqref{(1100)}, one gets
\begin{equation}\label{Eq.(1100)}
\begin{split}
&F^{1\bar{1}}+F^{2\bar{2}}-[\phi^{12},\phi^{34}]=0\,, \\
&F^{\bar{1}\bar{2}}+[\phi^{24},\phi^{34}]=0\,, \\
&D^{1}\phi^{34}-D^{\bar{2}}\phi^{13}=0\,, \\
&D^{1}\phi^{24}-D^{\bar{2}}\phi^{12}=0\,, \\
&[\phi^{23},\phi^{AB}]=[\phi^{14},\phi^{AB}]=
[\phi^{13}+\phi^{24},\phi^{AB}]=0\,, \\
&D^{\mu}\phi^{23}=D^{\mu}\phi^{14}=D^{\mu}(\phi^{13}+\phi^{24})=0\,.
\end{split}
\end{equation}
Therefore three purely imaginary fields decouple. Let us consider the
constraints coming from the existence of a third spinor of the same chirality.
The latter is completely generic and is of the form
\eqref{Generic Killing Spinor}-\eqref{eq: struttura simplettici}. Taking into
account \eqref{Eq.(1100)}, the system one obtains is
\begin{equation}\label{Eq.(2200)}
\begin{split}
&\mbox{Im}(d[\phi^{12},\phi^{24}])=0\,, \\
&\mbox{Im}(c[\phi^{12},\phi^{24}])=0\,, \\
&2 i \mbox{Im}(b) [\phi^{24},\phi^{34}] + c [\phi^{12},\phi^{34}]=0\,, \\
&2 i \mbox{Im}(a) [\phi^{24},\phi^{34}] + d [\phi^{12},\phi^{34}]=0\,, \\
&2 i \mbox{Im}(b) D^{\bar 2}\phi^{24} + c^* D^{\bar 2}\phi^{34}
-c D^1 \phi^{24}=0\,, \\
&2 i \mbox{Im}(a) D^{\bar 2}\phi^{24} + d^* D^{\bar 2}\phi^{34}
-d D^1 \phi^{24}=0\,, \\
&2 i \mbox{Im}(b) D^1\phi^{24} + c D^1 \phi^{12} +c^* D^{\bar 2}\phi^{24} =0\,, \\
&2 i \mbox{Im}(a) D^1\phi^{24} + d D^1 \phi^{12} +d^* D^{\bar 2}\phi^{24} =0\,. \\
\end{split}
\end{equation}
The precise form of the equations is not really important, but what is worth
noticing is that if this third spinor is indeed Killing, then it is also the
case for the two independent spinors on a given Lorentz orbit in which it can
be decomposed:
\begin{equation}\label{eq: scomposizione terzo}
  \xi_3=\left(
  \begin{array}{c}
    a\\b\\c\\d
  \end{array}
  \right)1+
  \left(
  \begin{array}{c}
    -d^*\\-c^*\\b^*\\a^*
  \end{array}
  \right)e_{12}=\left(
  \begin{array}{c}
    a\\0\\0\\d
  \end{array}
  \right)1+
  \left(
  \begin{array}{c}
    -d^*\\0\\0\\a^*
  \end{array}
  \right)e_{12}+\left(
  \begin{array}{c}
    0\\b\\c\\0
  \end{array}
  \right)1+
  \left(
  \begin{array}{c}
    0\\-c^*\\b^*\\0
  \end{array}
  \right)e_{12}\,,
\end{equation}
since the coefficients $(a,d)$ and $(b,c)$ do not mix. Furthermore, the
equations for the pair $(a,d)$ are exactly the same as the ones for $(b,c)$,
therefore if 
\begin{equation}
  \xi_3=\left(
  \begin{array}{c}
    a\\0\\0\\d
  \end{array}
  \right)1+
  \left(
  \begin{array}{c}
    -d^*\\0\\0\\a^*
  \end{array}
  \right)e_{12}
\end{equation}
is a Killing spinor, then automatically 
\begin{equation}\label{Right2ndOrbit}
  \xi_4=\left(
  \begin{array}{c}
    0\\a\\d\\0
  \end{array}
  \right)1+
  \left(
  \begin{array}{c}
    0\\-d^*\\a^*\\0
  \end{array}
  \right)e_{12}
\end{equation}
will also be Killing. Thus
\begin{center}
\scalebox{.9}{\begin{picture}(30,30)(30,0)
\multiframe(0,10)(10.5,0){1}(10,10){\footnotesize$\bullet$}
\multiframe(10,10)(10.5,0){1}(10,10){\footnotesize$\bullet$}
\multiframe(20,10)(10.5,0){1}(10,10){}
\multiframe(30,10)(10.5,0){1}(10,10){}

\multiframe(0,-0.5)(10.5,0){1}(10,10){\footnotesize$\bullet$}
\multiframe(10,-0.5)(10.5,0){1}(10,10){}
\multiframe(20,-0.5)(10.5,0){1}(10,10){}
\multiframe(30,-0.5)(10.5,0){1}(10,10){}
\end{picture}}
$\Rightarrow$
\scalebox{.9}{\begin{picture}(30,30)(-20,0)
\multiframe(0,10)(10.5,0){1}(10,10){\footnotesize$\bullet$}
\multiframe(10,10)(10.5,0){1}(10,10){\footnotesize$\bullet$}
\multiframe(20,10)(10.5,0){1}(10,10){}
\multiframe(30,10)(10.5,0){1}(10,10){}

\multiframe(0,-0.5)(10.5,0){1}(10,10){\footnotesize$\bullet$}
\multiframe(10,-0.5)(10.5,0){1}(10,10){\footnotesize$\bullet$}
\multiframe(20,-0.5)(10.5,0){1}(10,10){}
\multiframe(30,-0.5)(10.5,0){1}(10,10){}
\end{picture}}
\end{center}
For the sake of definiteness, let us choose $a=i$ and $d=0$. One then obtains
that a new scalar field, in particular $\phi^{24}$, decouples, and the equations
reduce to
\begin{equation}
\begin{split}
&F^{1\bar{1}}+F^{2\bar{2}}-[\phi^{12},\phi^{34}]=0\,, \\
&F^{\bar{1}\bar{2}}=0\,, \\
&D^{1}\phi^{34}=D^{2}\phi^{34} =0\,, \\
&D^{\bar{1}}\phi^{12}=D^{\bar{2}}\phi^{12}=0\,, \\
&[\phi^{23},\phi^{AB}]=[\phi^{14},\phi^{AB}]=[\phi^{13},\phi^{AB}]=
[\phi^{24},\phi^{AB}]=0\,, \\
&D^{\mu}\phi^{23}=D^{\mu}\phi^{14}=D^{\mu}\phi^{13}=D^{\mu}\phi^{24}=0\,.
\end{split}
\end{equation}
For a generic spinor, it is another combination of the scalar fields that
would decouple. Finally, adding one more right Killing spinor to the
$(2,2,0,0)$ configuration immediately leads to the instanton solution, i.e.~the
decoupling of all scalar fields and selfdual field strength, with eight
supersymmetries preserved:
\begin{center}
\scalebox{.9}{\begin{picture}(30,30)(30,0)
\multiframe(0,10)(10.5,0){1}(10,10){\footnotesize$\bullet$}
\multiframe(10,10)(10.5,0){1}(10,10){\footnotesize$\bullet$}
\multiframe(20,10)(10.5,0){1}(10,10){\footnotesize$\bullet$}
\multiframe(30,10)(10.5,0){1}(10,10){}

\multiframe(0,-0.5)(10.5,0){1}(10,10){\footnotesize$\bullet$}
\multiframe(10,-0.5)(10.5,0){1}(10,10){\footnotesize$\bullet$}
\multiframe(20,-0.5)(10.5,0){1}(10,10){}
\multiframe(30,-0.5)(10.5,0){1}(10,10){}
\end{picture}}
$\Rightarrow$
\scalebox{.9}{\begin{picture}(30,30)(-20,0)
\multiframe(0,10)(10.5,0){1}(10,10){\footnotesize$\bullet$}
\multiframe(10,10)(10.5,0){1}(10,10){\footnotesize$\bullet$}
\multiframe(20,10)(10.5,0){1}(10,10){\footnotesize$\bullet$}
\multiframe(30,10)(10.5,0){1}(10,10){\footnotesize$\bullet$}

\multiframe(0,-0.5)(10.5,0){1}(10,10){\footnotesize$\bullet$}
\multiframe(10,-0.5)(10.5,0){1}(10,10){\footnotesize$\bullet$}
\multiframe(20,-0.5)(10.5,0){1}(10,10){\footnotesize$\bullet$}
\multiframe(30,-0.5)(10.5,0){1}(10,10){\footnotesize$\bullet$}
\end{picture}}
\end{center}

The possible configurations with Killing spinors of the same chirality are
summarized in table \ref{Summary Chiral Killings}.

\begin{table}[hbt]

\begin{center}
\scalebox{.7}{\begin{picture}(30,30)(10,0)
\multiframe(0,10)(10.5,0){1}(10,10){\footnotesize$\bullet$}
\multiframe(10,10)(10.5,0){1}(10,10){}
\multiframe(20,10)(10.5,0){1}(10,10){}
\multiframe(30,10)(10.5,0){1}(10,10){}

\multiframe(0,-0.5)(10.5,0){1}(10,10){}
\multiframe(10,-0.5)(10.5,0){1}(10,10){}
\multiframe(20,-0.5)(10.5,0){1}(10,10){}
\multiframe(30,-0.5)(10.5,0){1}(10,10){}
\end{picture}} \qquad \mbox{One susy}
\end{center}

\begin{center}
\scalebox{.7}{\begin{picture}(30,30)(30,0)
\multiframe(0,10)(10.5,0){1}(10,10){\footnotesize$\bullet$}
\multiframe(10,10)(10.5,0){1}(10,10){\footnotesize$\bullet$}
\multiframe(20,10)(10.5,0){1}(10,10){}
\multiframe(30,10)(10.5,0){1}(10,10){}

\multiframe(0,-0.5)(10.5,0){1}(10,10){}
\multiframe(10,-0.5)(10.5,0){1}(10,10){}
\multiframe(20,-0.5)(10.5,0){1}(10,10){}
\multiframe(30,-0.5)(10.5,0){1}(10,10){}
\end{picture}}
\scalebox{.7}{\begin{picture}(30,30)(-20,0)
\multiframe(0,10)(10.5,0){1}(10,10){\footnotesize$\bullet$}
\multiframe(10,10)(10.5,0){1}(10,10){}
\multiframe(20,10)(10.5,0){1}(10,10){}
\multiframe(30,10)(10.5,0){1}(10,10){}

\multiframe(0,-0.5)(10.5,0){1}(10,10){\footnotesize$\bullet$}
\multiframe(10,-0.5)(10.5,0){1}(10,10){}
\multiframe(20,-0.5)(10.5,0){1}(10,10){}
\multiframe(30,-0.5)(10.5,0){1}(10,10){}
\end{picture}} \qquad \qquad \qquad \mbox{Two susys}
\end{center}

\begin{center}
\scalebox{.7}{\begin{picture}(30,30)(10,0)
\multiframe(0,10)(10.5,0){1}(10,10){\footnotesize$\bullet$}
\multiframe(10,10)(10.5,0){1}(10,10){\footnotesize$\bullet$}
\multiframe(20,10)(10.5,0){1}(10,10){\footnotesize$\bullet$}
\multiframe(30,10)(10.5,0){1}(10,10){}

\multiframe(0,-0.5)(10.5,0){1}(10,10){}
\multiframe(10,-0.5)(10.5,0){1}(10,10){}
\multiframe(20,-0.5)(10.5,0){1}(10,10){}
\multiframe(30,-0.5)(10.5,0){1}(10,10){}
\end{picture}} \qquad \mbox{Three susys  (F selfdual)}
\end{center}

\begin{center}
\scalebox{.7}{\begin{picture}(30,30)(10,0)
\multiframe(0,10)(10.5,0){1}(10,10){\footnotesize$\bullet$}
\multiframe(10,10)(10.5,0){1}(10,10){\footnotesize$\bullet$}
\multiframe(20,10)(10.5,0){1}(10,10){}
\multiframe(30,10)(10.5,0){1}(10,10){}

\multiframe(0,-0.5)(10.5,0){1}(10,10){\footnotesize$\bullet$}
\multiframe(10,-0.5)(10.5,0){1}(10,10){\footnotesize$\bullet$}
\multiframe(20,-0.5)(10.5,0){1}(10,10){}
\multiframe(30,-0.5)(10.5,0){1}(10,10){}
\end{picture}} \qquad \mbox{Four susys \qquad\qquad\quad}
\end{center}

\begin{center}
\scalebox{.7}{\begin{picture}(30,30)(10,0)
\multiframe(0,10)(10.5,0){1}(10,10){\footnotesize$\bullet$}
\multiframe(10,10)(10.5,0){1}(10,10){\footnotesize$\bullet$}
\multiframe(20,10)(10.5,0){1}(10,10){\footnotesize$\bullet$}
\multiframe(30,10)(10.5,0){1}(10,10){\footnotesize$\bullet$}

\multiframe(0,-0.5)(10.5,0){1}(10,10){\footnotesize$\bullet$}
\multiframe(10,-0.5)(10.5,0){1}(10,10){\footnotesize$\bullet$}
\multiframe(20,-0.5)(10.5,0){1}(10,10){\footnotesize$\bullet$}
\multiframe(30,-0.5)(10.5,0){1}(10,10){\footnotesize$\bullet$}
\end{picture}} \qquad \mbox{Eight susys: instanton}
\end{center}
\caption{\small{Possible supersymmetric configurations for Killing spinors of
the same chirality.}}\label{Summary Chiral Killings}
\end{table}

\subsubsection{Killing spinors with different chiralities}
We again start with the equations \eqref{Eq.FirstSpinor} imposed by the first
Killing spinor and from there we proceed methodically to take into
account the constraints coming from additional supersymmetries.
\newline
\newline
${\bf \underline{(1,0,0,0) \rightarrow (1,0,1,0)
\rightarrow (1,0,2,0) = (2,0,2,0)}}$
\newline
Combining the equations \eqref{Eq.FirstSpinor} from the first spinor with
those arising from plugging
\eqref{LeftChiral1} in \eqref{eq: vincoli geometria spinoriale}, one gets 
\begin{equation}\label{(1,0,1,0)} (1,0,1,0) \qquad \Leftrightarrow \qquad
\left\{\begin{array}{lll}
 F^{1\bar1} = [\phi^{12},\phi^{34}]+[\phi^{13} , \phi^{24}]\,, \\[.3em]
 F^{1\bar2} + [\phi^{12},\phi^{13}] =0\,, \qquad F^{2\bar 2}=0\,, \\[.3em]
 [\phi^{23},\phi^{AB}]=[\phi^{14},\phi^{AB}] = 0\quad\forall \; A,B\,, \\[.3em]
 D^\mu \phi^{23} = D^\mu \phi^{14}= 0\,, \\[.3em]
 [\phi^{12},\phi^{13}] = -[\phi^{12},\phi^{24}]\,, \\[.3em]
 [\phi^{13},\phi^{34}] = -[\phi^{24},\phi^{34}]\,, \\[.3em]
 (D^2 - D^{\bar 2})\phi^{AB} = 0\,. \end{array}\right .
\end{equation}
Therefore the field $\phi^{14}$ decouples, while subsequent Killing spinors
could never lead to a non-trivial selfdual solution. When adding a generic
left spinor \eqref{LeftChiralSpinor} to $(1,0,1,0)$, one sees similarly to
\eqref{Eq.(2200)} and \eqref{eq: scomposizione terzo} that the two spinors
lying on the Lorentz orbits of \eqref{LeftChiral1} and \eqref{LeftChiral2}
respectively into which it decomposes are also Killing, because the $(e,h)$
and $(f,g)$ components do not mix. Let us separate the two cases and continue
with $(1,0,2,0)$, thus implementing the constraints coming from 
\begin{equation}\label{Kill(1020)}
  \bar\xi_3=\left(
  \begin{array}{c}
    e\\0\\0\\h
  \end{array}
  \right)e_1+
  \left(
  \begin{array}{c}
    h^*\\0\\0\\-e^*
  \end{array}
  \right)e_{2}\ .
\end{equation}
One verifies that these constraints automatically imply that 
\begin{equation}\label{4thKS}
  \xi_4=\left(
  \begin{array}{c}
    e\\0\\0\\-h^*
  \end{array}
  \right)1+
  \left(
  \begin{array}{c}
    h\\0\\0\\e^*
  \end{array}
  \right)e_{12}
\end{equation}
belonging to the Lorentz orbit of \eqref{eq: rappresentativo chirale} is a
Killing spinor, therefore
\begin{table}[h]
\begin{center}
\scalebox{.9}{\begin{picture}(30,30)(30,0)
\multiframe(0,10)(10.5,0){1}(10,10){\footnotesize$\bullet$}
\multiframe(10,10)(10.5,0){1}(10,10){}
\multiframe(20,10)(10.5,0){1}(10,10){}
\multiframe(30,10)(10.5,0){1}(10,10){}

\multiframe(0,-0.5)(10.5,0){1}(10,10){}
\multiframe(10,-0.5)(10.5,0){1}(10,10){}
\multiframe(20,-0.5)(10.5,0){1}(10,10){}
\multiframe(30,-0.5)(10.5,0){1}(10,10){}

\multiframe(0,-11)(10.5,0){1}(10,10){\footnotesize$\bullet$}
\multiframe(10,-11)(10.5,0){1}(10,10){\footnotesize$\bullet$}
\multiframe(20,-11)(10.5,0){1}(10,10){}
\multiframe(30,-11)(10.5,0){1}(10,10){}

\multiframe(0,-21.5)(10.5,0){1}(10,10){}
\multiframe(10,-21.5)(10.5,0){1}(10,10){}
\multiframe(20,-21.5)(10.5,0){1}(10,10){}
\multiframe(30,-21.5)(10.5,0){1}(10,10){}
\end{picture}}
$\Rightarrow$
\scalebox{.9}{\begin{picture}(30,30)(-20,0)
\multiframe(0,10)(10.5,0){1}(10,10){\footnotesize$\bullet$}
\multiframe(10,10)(10.5,0){1}(10,10){\footnotesize$\bullet$}
\multiframe(20,10)(10.5,0){1}(10,10){}
\multiframe(30,10)(10.5,0){1}(10,10){}

\multiframe(0,-0.5)(10.5,0){1}(10,10){}
\multiframe(10,-0.5)(10.5,0){1}(10,10){}
\multiframe(20,-0.5)(10.5,0){1}(10,10){}
\multiframe(30,-0.5)(10.5,0){1}(10,10){}

\multiframe(0,-11)(10.5,0){1}(10,10){\footnotesize$\bullet$}
\multiframe(10,-11)(10.5,0){1}(10,10){\footnotesize$\bullet$}
\multiframe(20,-11)(10.5,0){1}(10,10){}
\multiframe(30,-11)(10.5,0){1}(10,10){}

\multiframe(0,-21.5)(10.5,0){1}(10,10){}
\multiframe(10,-21.5)(10.5,0){1}(10,10){}
\multiframe(20,-21.5)(10.5,0){1}(10,10){}
\multiframe(30,-21.5)(10.5,0){1}(10,10){}
\end{picture}}
\end{center}
\end{table}
\newline
\newline
${\bf\underline{(1,0,0,0)\rightarrow (1,0,1,0)\rightarrow (1,0,1,1) =
(1,1,1,1) \rightarrow (2,1,1,1)=(2,2,2,2)\rightarrow (3,2,2,2)}}$\\
${\bf\underline{=(4,4,4,4)}}$
\newline
Starting from $(1,0,1,0)$, one would like to combine the constraints
\eqref{(1,0,1,0)} with those coming from a spinor of the form
\begin{equation}\label{Left2ndOrbit}
  \bar\xi_3=
  \left(
  \begin{array}{c}
    0\\f\\g\\0
  \end{array}
  \right)e_1+
  \left(
  \begin{array}{c}
    0\\g^*\\-f^*\\0
  \end{array}
  \right)e_2\,.
\end{equation}
Actually, this form can be simplified using the subgroup stabilizing
\eqref{eq: rappresentativo chirale} and \eqref{LeftChiral1}, composed of the 6
generators $S_1,\ldots,S_3$ and $S_8,\ldots,S_{10}$. (The latter, of course,
do not act on $\bar\xi_3$). With $f=B\exp(2i\varphi_B)$ and
$g=C\exp(2i\varphi_C)$, $\bar\xi_3$ can be brought to 
\begin{equation}
  \bar\xi_3=
  \left(
  \begin{array}{c}
    0\\B \sqrt{1 + C^2/B^2}\\0\\0
  \end{array}
  \right)e_1+
  \left(
  \begin{array}{c}
    0\\0\\-B \sqrt{1 + C^2/B^2}\\0
  \end{array}
  \right)e_2
  \end{equation}
by acting with $\exp[(\varphi_B+\varphi_C)S_1]
\cdot\exp[(\mbox{arctg}\,C/B)S_2]\cdot\exp[(\varphi_B-\varphi_C)S_1]$.
Equations \eqref{(1,0,1,0)} are then supplemented by the conditions
\begin{equation}\label{(1011)}
\begin{split}
&D^{1}\phi^{34}+D^{\bar{2}}\phi^{24}=0\,, \\
&D^{\bar1}\phi^{24}-D^{2}\phi^{34}=0\,, \\
&D^1 (\phi^{13} + \phi^{24})=0\,, \\
&[\phi^{13},\phi^{24}]=0\,. \\
\end{split}
\end{equation}
One may then check that the susy equations for the Killing spinor \eqref{(1100)}
are satisfied as a consequence of \eqref{(1,0,1,0)} and \eqref{(1011)}, yielding
\begin{table}[h]
\begin{center}
\scalebox{.9}{\begin{picture}(30,30)(30,0)
\multiframe(0,10)(10.5,0){1}(10,10){\footnotesize$\bullet$}
\multiframe(10,10)(10.5,0){1}(10,10){}
\multiframe(20,10)(10.5,0){1}(10,10){}
\multiframe(30,10)(10.5,0){1}(10,10){}

\multiframe(0,-0.5)(10.5,0){1}(10,10){}
\multiframe(10,-0.5)(10.5,0){1}(10,10){}
\multiframe(20,-0.5)(10.5,0){1}(10,10){}
\multiframe(30,-0.5)(10.5,0){1}(10,10){}

\multiframe(0,-11)(10.5,0){1}(10,10){\footnotesize$\bullet$}
\multiframe(10,-11)(10.5,0){1}(10,10){}
\multiframe(20,-11)(10.5,0){1}(10,10){}
\multiframe(30,-11)(10.5,0){1}(10,10){}

\multiframe(0,-21.5)(10.5,0){1}(10,10){\footnotesize$\bullet$}
\multiframe(10,-21.5)(10.5,0){1}(10,10){}
\multiframe(20,-21.5)(10.5,0){1}(10,10){}
\multiframe(30,-21.5)(10.5,0){1}(10,10){}
\end{picture}}
$\Rightarrow$
\scalebox{.9}{\begin{picture}(30,30)(-20,0)
\multiframe(0,10)(10.5,0){1}(10,10){\footnotesize$\bullet$}
\multiframe(10,10)(10.5,0){1}(10,10){}
\multiframe(20,10)(10.5,0){1}(10,10){}
\multiframe(30,10)(10.5,0){1}(10,10){}

\multiframe(0,-0.5)(10.5,0){1}(10,10){\footnotesize$\bullet$}
\multiframe(10,-0.5)(10.5,0){1}(10,10){}
\multiframe(20,-0.5)(10.5,0){1}(10,10){}
\multiframe(30,-0.5)(10.5,0){1}(10,10){}

\multiframe(0,-11)(10.5,0){1}(10,10){\footnotesize$\bullet$}
\multiframe(10,-11)(10.5,0){1}(10,10){}
\multiframe(20,-11)(10.5,0){1}(10,10){}
\multiframe(30,-11)(10.5,0){1}(10,10){}

\multiframe(0,-21.5)(10.5,0){1}(10,10){\footnotesize$\bullet$}
\multiframe(10,-21.5)(10.5,0){1}(10,10){}
\multiframe(20,-21.5)(10.5,0){1}(10,10){}
\multiframe(30,-21.5)(10.5,0){1}(10,10){}
\end{picture}}
\end{center}
\end{table}
\newline
\newline
As a next step, let us impose an additional Killing spinor in one of the
Lorentz orbits, say the first without loss of generality. We thus take it of
the form \eqref{(2000)}, in such a way that it be linearly independent of the
first one. The equations coming out by combining the new constraints with the
former ones are not really enlightening, so we just mention the expression of
the field strength:
\begin{eqnarray}
&&F^{1\bar{1}}=\frac{2i}{d^*}\mbox{Im}(a)[\phi^{12},\phi^{24}]\ , \qquad
F^{2\bar{2}}=0\ , \\
&&F^{\bar{1}\bar{2}}=F^{\bar{1}2}=\frac{d}{d^*} [\phi^{12},\phi^{24}]\ . \nonumber
\end{eqnarray}
However, more importantly, one can check that after having imposed the
conditions for the latter $(2,1,1,1)$ configuration, the following spinors are
Killing: \eqref{Right2ndOrbit} with $a=i$, \eqref{Kill(1020)} with
$h=-d^* \mbox{Im}(e)$ and \eqref{Left2ndOrbit} with $g=-d^* \mbox{Im}(f)$, and
hence
\begin{table}[h]
\begin{center}
\scalebox{.9}{\begin{picture}(30,30)(30,0)
\multiframe(0,10)(10.5,0){1}(10,10){\footnotesize$\bullet$}
\multiframe(10,10)(10.5,0){1}(10,10){\footnotesize$\bullet$}
\multiframe(20,10)(10.5,0){1}(10,10){}
\multiframe(30,10)(10.5,0){1}(10,10){}

\multiframe(0,-0.5)(10.5,0){1}(10,10){\footnotesize$\bullet$}
\multiframe(10,-0.5)(10.5,0){1}(10,10){}
\multiframe(20,-0.5)(10.5,0){1}(10,10){}
\multiframe(30,-0.5)(10.5,0){1}(10,10){}

\multiframe(0,-11)(10.5,0){1}(10,10){\footnotesize$\bullet$}
\multiframe(10,-11)(10.5,0){1}(10,10){}
\multiframe(20,-11)(10.5,0){1}(10,10){}
\multiframe(30,-11)(10.5,0){1}(10,10){}

\multiframe(0,-21.5)(10.5,0){1}(10,10){\footnotesize$\bullet$}
\multiframe(10,-21.5)(10.5,0){1}(10,10){}
\multiframe(20,-21.5)(10.5,0){1}(10,10){}
\multiframe(30,-21.5)(10.5,0){1}(10,10){}
\end{picture}}
$\Rightarrow$
\scalebox{.9}{\begin{picture}(30,30)(-20,0)
\multiframe(0,10)(10.5,0){1}(10,10){\footnotesize$\bullet$}
\multiframe(10,10)(10.5,0){1}(10,10){\footnotesize$\bullet$}
\multiframe(20,10)(10.5,0){1}(10,10){}
\multiframe(30,10)(10.5,0){1}(10,10){}

\multiframe(0,-0.5)(10.5,0){1}(10,10){\footnotesize$\bullet$}
\multiframe(10,-0.5)(10.5,0){1}(10,10){\footnotesize$\bullet$}
\multiframe(20,-0.5)(10.5,0){1}(10,10){}
\multiframe(30,-0.5)(10.5,0){1}(10,10){}

\multiframe(0,-11)(10.5,0){1}(10,10){\footnotesize$\bullet$}
\multiframe(10,-11)(10.5,0){1}(10,10){\footnotesize$\bullet$}
\multiframe(20,-11)(10.5,0){1}(10,10){}
\multiframe(30,-11)(10.5,0){1}(10,10){}

\multiframe(0,-21.5)(10.5,0){1}(10,10){\footnotesize$\bullet$}
\multiframe(10,-21.5)(10.5,0){1}(10,10){\footnotesize$\bullet$}
\multiframe(20,-21.5)(10.5,0){1}(10,10){}
\multiframe(30,-21.5)(10.5,0){1}(10,10){}
\end{picture}}
\end{center}
\end{table}
\newline
\newline
At last, adding the constraints of one more Killing spinor belonging to one of
the four Lorentz orbits, say the first, leads to the vacuum solution preserving
all 16 supersymmetries:
\begin{table}[h]
\begin{center}
\scalebox{.9}{\begin{picture}(30,30)(30,0)
\multiframe(0,10)(10.5,0){1}(10,10){\footnotesize$\bullet$}
\multiframe(10,10)(10.5,0){1}(10,10){\footnotesize$\bullet$}
\multiframe(20,10)(10.5,0){1}(10,10){\footnotesize$\bullet$}
\multiframe(30,10)(10.5,0){1}(10,10){}

\multiframe(0,-0.5)(10.5,0){1}(10,10){\footnotesize$\bullet$}
\multiframe(10,-0.5)(10.5,0){1}(10,10){\footnotesize$\bullet$}
\multiframe(20,-0.5)(10.5,0){1}(10,10){}
\multiframe(30,-0.5)(10.5,0){1}(10,10){}

\multiframe(0,-11)(10.5,0){1}(10,10){\footnotesize$\bullet$}
\multiframe(10,-11)(10.5,0){1}(10,10){\footnotesize$\bullet$}
\multiframe(20,-11)(10.5,0){1}(10,10){}
\multiframe(30,-11)(10.5,0){1}(10,10){}

\multiframe(0,-21.5)(10.5,0){1}(10,10){\footnotesize$\bullet$}
\multiframe(10,-21.5)(10.5,0){1}(10,10){\footnotesize$\bullet$}
\multiframe(20,-21.5)(10.5,0){1}(10,10){}
\multiframe(30,-21.5)(10.5,0){1}(10,10){}
\end{picture}}
$\Rightarrow$
\scalebox{.9}{\begin{picture}(30,30)(-20,0)
\multiframe(0,10)(10.5,0){1}(10,10){\footnotesize$\bullet$}
\multiframe(10,10)(10.5,0){1}(10,10){\footnotesize$\bullet$}
\multiframe(20,10)(10.5,0){1}(10,10){\footnotesize$\bullet$}
\multiframe(30,10)(10.5,0){1}(10,10){\footnotesize$\bullet$}

\multiframe(0,-0.5)(10.5,0){1}(10,10){\footnotesize$\bullet$}
\multiframe(10,-0.5)(10.5,0){1}(10,10){\footnotesize$\bullet$}
\multiframe(20,-0.5)(10.5,0){1}(10,10){\footnotesize$\bullet$}
\multiframe(30,-0.5)(10.5,0){1}(10,10){\footnotesize$\bullet$}

\multiframe(0,-11)(10.5,0){1}(10,10){\footnotesize$\bullet$}
\multiframe(10,-11)(10.5,0){1}(10,10){\footnotesize$\bullet$}
\multiframe(20,-11)(10.5,0){1}(10,10){\footnotesize$\bullet$}
\multiframe(30,-11)(10.5,0){1}(10,10){\footnotesize$\bullet$}

\multiframe(0,-21.5)(10.5,0){1}(10,10){\footnotesize$\bullet$}
\multiframe(10,-21.5)(10.5,0){1}(10,10){\footnotesize$\bullet$}
\multiframe(20,-21.5)(10.5,0){1}(10,10){\footnotesize$\bullet$}
\multiframe(30,-21.5)(10.5,0){1}(10,10){\footnotesize$\bullet$}
\end{picture}}
\end{center}
\end{table}
\newline
\newline
\quad \\
${\bf\underline{(2,0,2,0)\rightarrow (2,0,3,0) = (4,0,4,0)\rightarrow
(4,0,4,1) = (4,4,4,4) }}$
\newline
Supplementing the constraints of $(2,0,2,0)$ by the one originating from one
more left Killing spinor on the same Lorentz orbit, configuration $(2,0,3,0)$
leads to the vanishing of the field strength, all covariant derivatives and all
commutators but two, $[\phi^{12},\phi^{34}] = [\phi^{13},\phi^{24}]$.
These restrictions imply the supersymmetry equations for any spinor on the
orbits of \eqref{eq: rappresentativo chirale} or \eqref{LeftChiral1}, and
consequently
\begin{table}[h]
\begin{center}
\scalebox{.8}{\begin{picture}(30,30)(30,0)
\multiframe(0,10)(10.5,0){1}(10,10){\footnotesize$\bullet$}
\multiframe(10,10)(10.5,0){1}(10,10){\footnotesize$\bullet$}
\multiframe(20,10)(10.5,0){1}(10,10){}
\multiframe(30,10)(10.5,0){1}(10,10){}

\multiframe(0,-0.5)(10.5,0){1}(10,10){}
\multiframe(10,-0.5)(10.5,0){1}(10,10){}
\multiframe(20,-0.5)(10.5,0){1}(10,10){}
\multiframe(30,-0.5)(10.5,0){1}(10,10){}

\multiframe(0,-11)(10.5,0){1}(10,10){\footnotesize$\bullet$}
\multiframe(10,-11)(10.5,0){1}(10,10){\footnotesize$\bullet$}
\multiframe(20,-11)(10.5,0){1}(10,10){\footnotesize$\bullet$}
\multiframe(30,-11)(10.5,0){1}(10,10){}

\multiframe(0,-21.5)(10.5,0){1}(10,10){}
\multiframe(10,-21.5)(10.5,0){1}(10,10){}
\multiframe(20,-21.5)(10.5,0){1}(10,10){}
\multiframe(30,-21.5)(10.5,0){1}(10,10){}
\end{picture}}
$\Rightarrow$
\scalebox{.8}{\begin{picture}(30,30)(-20,0)
\multiframe(0,10)(10.5,0){1}(10,10){\footnotesize$\bullet$}
\multiframe(10,10)(10.5,0){1}(10,10){\footnotesize$\bullet$}
\multiframe(20,10)(10.5,0){1}(10,10){\footnotesize$\bullet$}
\multiframe(30,10)(10.5,0){1}(10,10){\footnotesize$\bullet$}

\multiframe(0,-0.5)(10.5,0){1}(10,10){}
\multiframe(10,-0.5)(10.5,0){1}(10,10){}
\multiframe(20,-0.5)(10.5,0){1}(10,10){}
\multiframe(30,-0.5)(10.5,0){1}(10,10){}

\multiframe(0,-11)(10.5,0){1}(10,10){\footnotesize$\bullet$}
\multiframe(10,-11)(10.5,0){1}(10,10){\footnotesize$\bullet$}
\multiframe(20,-11)(10.5,0){1}(10,10){\footnotesize$\bullet$}
\multiframe(30,-11)(10.5,0){1}(10,10){\footnotesize$\bullet$}

\multiframe(0,-21.5)(10.5,0){1}(10,10){}
\multiframe(10,-21.5)(10.5,0){1}(10,10){}
\multiframe(20,-21.5)(10.5,0){1}(10,10){}
\multiframe(30,-21.5)(10.5,0){1}(10,10){}
\end{picture}}
\end{center}
\end{table}
\newline  
Finally, adding any spinor to $(4,0,4,0)$ leads to the vanishing of the last
commutators, and thus to the vacuum solution preserving all 16
supersymmetries:
\begin{table}[h]
\begin{center}
\scalebox{.9}{\begin{picture}(30,30)(30,0)
\multiframe(0,10)(10.5,0){1}(10,10){\footnotesize$\bullet$}
\multiframe(10,10)(10.5,0){1}(10,10){\footnotesize$\bullet$}
\multiframe(20,10)(10.5,0){1}(10,10){\footnotesize$\bullet$}
\multiframe(30,10)(10.5,0){1}(10,10){\footnotesize$\bullet$}

\multiframe(0,-0.5)(10.5,0){1}(10,10){}
\multiframe(10,-0.5)(10.5,0){1}(10,10){}
\multiframe(20,-0.5)(10.5,0){1}(10,10){}
\multiframe(30,-0.5)(10.5,0){1}(10,10){}

\multiframe(0,-11)(10.5,0){1}(10,10){\footnotesize$\bullet$}
\multiframe(10,-11)(10.5,0){1}(10,10){\footnotesize$\bullet$}
\multiframe(20,-11)(10.5,0){1}(10,10){\footnotesize$\bullet$}
\multiframe(30,-11)(10.5,0){1}(10,10){\footnotesize$\bullet$}

\multiframe(0,-21.5)(10.5,0){1}(10,10){\footnotesize$\bullet$}
\multiframe(10,-21.5)(10.5,0){1}(10,10){}
\multiframe(20,-21.5)(10.5,0){1}(10,10){}
\multiframe(30,-21.5)(10.5,0){1}(10,10){}
\end{picture}}
$\Rightarrow$
\scalebox{.9}{\begin{picture}(30,30)(-20,0)
\multiframe(0,10)(10.5,0){1}(10,10){\footnotesize$\bullet$}
\multiframe(10,10)(10.5,0){1}(10,10){\footnotesize$\bullet$}
\multiframe(20,10)(10.5,0){1}(10,10){\footnotesize$\bullet$}
\multiframe(30,10)(10.5,0){1}(10,10){\footnotesize$\bullet$}

\multiframe(0,-0.5)(10.5,0){1}(10,10){\footnotesize$\bullet$}
\multiframe(10,-0.5)(10.5,0){1}(10,10){\footnotesize$\bullet$}
\multiframe(20,-0.5)(10.5,0){1}(10,10){\footnotesize$\bullet$}
\multiframe(30,-0.5)(10.5,0){1}(10,10){\footnotesize$\bullet$}

\multiframe(0,-11)(10.5,0){1}(10,10){\footnotesize$\bullet$}
\multiframe(10,-11)(10.5,0){1}(10,10){\footnotesize$\bullet$}
\multiframe(20,-11)(10.5,0){1}(10,10){\footnotesize$\bullet$}
\multiframe(30,-11)(10.5,0){1}(10,10){\footnotesize$\bullet$}

\multiframe(0,-21.5)(10.5,0){1}(10,10){\footnotesize$\bullet$}
\multiframe(10,-21.5)(10.5,0){1}(10,10){\footnotesize$\bullet$}
\multiframe(20,-21.5)(10.5,0){1}(10,10){\footnotesize$\bullet$}
\multiframe(30,-21.5)(10.5,0){1}(10,10){\footnotesize$\bullet$}
\end{picture}}
\end{center}
\end{table}
\newline  
\newline
${\bf \underline{(1,0,0,0)\rightarrow(1,0,0,1)\rightarrow (1,0,0,2)
\rightarrow (1,0,0,3)\rightarrow (1,0,0,4)\rightarrow (2,0,0,4)}}$
\newline
Let us now re-start with the second Killing spinor of opposite chirality on the
other orbit, thus of the form \eqref{LeftChiral2}. Combining its constraints
with those of the first Killing spinor yields for the field strength
\begin{eqnarray}
&&F^{1\bar{1}}=[\phi^{12},\phi^{34}]\,, \qquad F^{2\bar{2}}=-[\phi^{13},\phi^{24}]
\,, \label{Eq.1001} \\
&&F^{\bar{1}\bar{2}}=-[\phi^{24},\phi^{34}]\,, \qquad F^{\bar{1}2}=[\phi^{13},
\phi^{34}]\,, \nonumber
\end{eqnarray}
or equivalently
\begin{equation}
F - \phi\wedge\phi = 0\,.
\end{equation}
One can subsequently look for the existence of additional Killing spinors
belonging for example to the Lorentz orbit of the second Killing spinor.
Naturally, after the second one, all spinors cannot be further simplified and
have to be taken generic, of the form \eqref{Left2ndOrbit}. Again, the equations
involving the covariant derivatives of the scalar fields are rather involved
and do not tell much, so we will focus on the field strength. From $(1,0,0,2)$,
one observes that $F^{\bar1 2}=0$, although F is not yet anti-selfdual. For the
configuration $(1,0,0,3)$, as expected from section \ref{same}, the field
strength becomes anti-selfdual (as the for the configuration $(0,0,0,3)$),
while the remaining component $F^{\bar{1}\bar{2}}$ is now determined in terms of
the commutator of scalars:
\begin{eqnarray}
&&F^{1\bar{1}} = F^{2\bar{2}}= -[\phi^{13},\phi^{24}]\,, \qquad F^{\bar{1}2} = 0
\,, \label{Eq.1003} \\
&&F^{\bar{1}\bar{2}}=-[\phi^{24},\phi^{34}]\,. \nonumber
\end{eqnarray}
For $(1,0,0,4)$, we get of course something very similar to \eqref{4000}, with
the difference that the field strength is anti-selfdual instead of selfdual,
and with the additional information that can be seen in \eqref{Eq.1003} that
the components of $F$ are related to commutators of scalar fields (which was
not the case in \eqref{4000}). Adding one more Killing spinor on the orbit of
the first spinor further yields $F^{\bar{1}\bar{2}}=-[\phi^{24},\phi^{34}]=0$.
\newline
\newline
${\bf \underline{(1,0,0,2)\rightarrow (2,0,0,2)\rightarrow (3,0,0,2)/(2,0,0,3)
\rightarrow (3,0,0,3)\rightarrow (4,0,0,3) = (4,0,0,4)}}$
\newline
The last cases that haven't yet been explored or that are not a consequence of
what we have seen up to now consist in adding to $(1,0,0,2)$ a Killing spinor
in the Lorentz orbit of the first representative. The configuration $(2,0,0,2)$
will not be (anti-)selfdual, but has $F^{\bar1 2}= F^{12}=0$. Next, as expected,
$(3,0,0,2)$ is selfdual while $(2,0,0,3)$ is anti-selfdual. As a consequence,
$(3,0,0,3)$ has vanishing field strength. Finally, one can see using a basis
for elements on the first and fourth orbit that 
\begin{table}[h]
\begin{center}
\scalebox{.9}{\begin{picture}(30,30)(30,0)
\multiframe(0,10)(10.5,0){1}(10,10){\footnotesize$\bullet$}
\multiframe(10,10)(10.5,0){1}(10,10){\footnotesize$\bullet$}
\multiframe(20,10)(10.5,0){1}(10,10){\footnotesize$\bullet$}
\multiframe(30,10)(10.5,0){1}(10,10){\footnotesize$\bullet$}

\multiframe(0,-0.5)(10.5,0){1}(10,10){}
\multiframe(10,-0.5)(10.5,0){1}(10,10){}
\multiframe(20,-0.5)(10.5,0){1}(10,10){}
\multiframe(30,-0.5)(10.5,0){1}(10,10){}

\multiframe(0,-11)(10.5,0){1}(10,10){}
\multiframe(10,-11)(10.5,0){1}(10,10){}
\multiframe(20,-11)(10.5,0){1}(10,10){}
\multiframe(30,-11)(10.5,0){1}(10,10){}

\multiframe(0,-21.5)(10.5,0){1}(10,10){\footnotesize$\bullet$}
\multiframe(10,-21.5)(10.5,0){1}(10,10){\footnotesize$\bullet$}
\multiframe(20,-21.5)(10.5,0){1}(10,10){\footnotesize$\bullet$}
\multiframe(30,-21.5)(10.5,0){1}(10,10){}
\end{picture}}
$\Rightarrow$
\scalebox{.9}{\begin{picture}(30,30)(-20,0)
\multiframe(0,10)(10.5,0){1}(10,10){\footnotesize$\bullet$}
\multiframe(10,10)(10.5,0){1}(10,10){\footnotesize$\bullet$}
\multiframe(20,10)(10.5,0){1}(10,10){\footnotesize$\bullet$}
\multiframe(30,10)(10.5,0){1}(10,10){\footnotesize$\bullet$}

\multiframe(0,-0.5)(10.5,0){1}(10,10){}
\multiframe(10,-0.5)(10.5,0){1}(10,10){}
\multiframe(20,-0.5)(10.5,0){1}(10,10){}
\multiframe(30,-0.5)(10.5,0){1}(10,10){}

\multiframe(0,-11)(10.5,0){1}(10,10){}
\multiframe(10,-11)(10.5,0){1}(10,10){}
\multiframe(20,-11)(10.5,0){1}(10,10){}
\multiframe(30,-11)(10.5,0){1}(10,10){}

\multiframe(0,-21.5)(10.5,0){1}(10,10){\footnotesize$\bullet$}
\multiframe(10,-21.5)(10.5,0){1}(10,10){\footnotesize$\bullet$}
\multiframe(20,-21.5)(10.5,0){1}(10,10){\footnotesize$\bullet$}
\multiframe(30,-21.5)(10.5,0){1}(10,10){\footnotesize$\bullet$}
\end{picture}}
\end{center}
\end{table}

The classification of supersymmetric backgrounds of ${\cal N}=4$ SYM theory
is summarized in table \ref{Recap}.

\begin{table}[ht]
\begin{center}
\begin{tabular}{||c|c|c|c|c|c|c|c||}
\hline 1 & 2 & 3 & 4 &5 & 6 & 8 & 16\\
\hline
(1,0,0,0) & (2,0,0,0) & (3,0,0,0) & (4,0,0,0) & (3,0,0,2) & (3,0,0,3) &
(4,4,0,0) & (4,4,4,4) \\
& (1,1,0,0) & (2,0,0,1) & (2,2,0,0) & (4,0,0,1) & (4,0,0,2) & (4,0,4,0) &
\; \\
& (1,0,1,0) & & (2,0,2,0) & & & (2,2,2,2) & \; \\             
& (1,0,0,1) & & (1,1,1,1) & & & (4,0,0,4) & \; \\
&           & & (2,0,0,2) & & &           & \; \\
&           & & (3,0,0,1) & & &           & \;\\
\hline
\end{tabular}
\end{center}
\caption{Classification of purely bosonic supersymmetric configurations of
${\cal N}=4$ SYM theory, with Killing spinors of definite chiralities. The first line indicates the number $n$ of supersymmetries. Notice that there are no backgrounds with $n=7$ or
$9\le n\le15$. An arbitrary configuration $(n_1,n_2,n_3,n_4)$ if not present in the table can be shown to be equivalent to one of the above using the analysis of the previous section. As an example, let us consider $(2,0,2,1)$. We have seen that  $(1,0,1,1)=(1,1,1,1)$ (the equality sign meaning "implies"), therefore $(2,0,2,1)=(2,1,2,1)$. But we also observed that $(2,1,1,1) = (2,2,2,2)$. Therefore $(2,0,2,1)=(2,2,2,2)$.}\label{Recap}
\end{table}

\section{Susy variations and equations of motion}
\label{susyeom}

The Killing spinor equations arising from setting to zero the supersymmetry
variations \eqref{SusyEqCompl} are first order, and one could ask whether
they imply the second order equations of motion
\eqref{eomF}-\eqref{eombarlambda}. In
supergravity, this is not always the case: The Killing vector constructed as a
bilinear from
the Killing spinor can be either timelike or lightlike. One can show that in
the former case, the Killing spinor equations, together with the Bianchi
identities and the Maxwell equations, do entail the Einstein equations,
whereas in the null case, one of the Einstein equations must be additionally
imposed by hand (cf.~e.g.~\cite{Gauntlett:2002nw}).

We first focus on purely bosonic configurations. Let us consider the conditions
\eqref{Eq.FirstSpinor} imposed by the existence of a chiral Killing spinor. One
would like to check whether these imply
\begin{equation}\label{eq: motoBoso}
  \left\{
  \begin{split}
    &D^\nu F_{\nu\mu}
      -\frac{1}{2}[\bar\phi_{AB},D_{\mu}\phi^{AB}]=0\ ,\\
    &D^2\phi^{AB}
      -\frac{1}{2}[\bar\phi_{CD},[\phi^{AB},\phi^{CD}]]=0\ .\\
      \end{split}
  \right.
\end{equation}
A first observation is that one the scalar fields, here $\phi^{14}$, does not
appear in the susy equations (except through the fact that it commutes with
$\phi^{23}$). Therefore its equations of motion are certainly not satisfied by
virtue of the susy equations, and the former will have to be imposed by hand.
However, all the other equations of motion will automatically hold, using only
\eqref{Eq.FirstSpinor}, the Jacobi identity and the Bianchi identities for the
gauge field, as we illustrate now.

Consider the first equation of \eqref{eq: motoBoso} for e.g.~$\mu=1$. By
expanding it, using the definition
$\bar\phi_{AB}=\frac12\epsilon_{ABCD}\phi^{CD}$ for the duals and the fact
that $\phi^{23}$ decouples, one gets
\begin{eqnarray}
\lefteqn{D^{\bar 1}F^{1\bar 1} + D^{2}F^{\bar 2\bar 1}+D^{\bar 2}F^{2\bar 1}}
\nonumber \\
&&- [\phi^{34},D_1 \phi^{12}] - [\phi^{42},D_1 \phi^{13}] - [\phi^{31},D_1
\phi^{24}] - [\phi^{12},D_1 \phi^{34}] \overset{?}{=} 0\ . \label{EOM1OK}
\end{eqnarray}
Now use the Bianchi identity
\begin{equation}
 D^{\bar 2}F^{2\bar 1} = D^{2}F^{\bar 2\bar 1} +D^{\bar 1}F^{2\bar 2}\ ,
\end{equation}
to replace the term
$D^{\bar 1}F^{1\bar 1}+D^{2}F^{\bar 2\bar 1}+D^{\bar 2}F^{2\bar 1}$ in
\eqref{EOM1OK} by $2D^{2}F^{\bar 2\bar 1}+D^{\bar 1}(F^{1\bar 1}+F^{2\bar 2})$ and
finally the two first equations of \eqref{Eq.FirstSpinor} to eliminate the
field strengths in terms of commutators of scalar fields, to see that they
exactly cancel the scalar field part in \eqref{EOM1OK}. The other components
are satisfied in the same way.

We now turn to the second equation of \eqref{eq: motoBoso} and show explicitly
how things combine for $A=1$, $B=3$. (For all other index combinations except
$A=1$, $B=4$, the proof is analogous).
First the $D_\mu D^\mu\phi^{13}$ term is
rewritten as
$2(D^1 D^{\bar 1}+D^2 D^{\bar 2})\phi^{13}-[F^{1\bar 1}+F^{2\bar 2},\phi^{13}]$
which, using the third equation of \eqref{Eq.FirstSpinor}, as well as the
complex conjugate of the fourth, boils down to
$-2[F^{12},\phi^{13}]-[F^{1\bar 1}+F^{2\bar 2},\phi^{13}]$. Using then the first
and the complex conjugate of the second equation of \eqref{Eq.FirstSpinor}, one
arrives at a sum of 3 multicommutator terms involving only
$\phi^{12}$, $\phi^{13}$ and $\phi^{34}$ which vanishes by virtue of the Jacobi
identity.

\section{Final remarks}
\label{final}

In this work we have discussed the classification of supersymmetric solutions
of euclidean ${\cal N}=4$ SYM theory in 4 dimensions. We have displayed the
equations they satisfy in the spinorial geometry language. The equations one
gets by imposing the existence of a single Killing spinor can be obtained by
dimensional reduction from 8 to 4 dimensions of the octonionic instanton
equations, much like the Hitchin equations arise by reduction of the
selfduality equations from 4 to 2 dimensions. The system of equations may be
rephrased compactly in terms of a selfduality-like condition for a complex
connection. We next wrote down the equations arising by imposing the existence
of more Killing spinors (focusing on Killing spinors with definite chiralities). Their analysis led to the conclusion that not all
fractions of the maximal number of supersymmetries are allowed. In particular,
there are no bosonic configurations preserving 7 supersymmetries (because this
automatically implies the existence of an 8th Killing spinor) nor 9 to 15 (that
would imply 16).  This is reminiscent of more familiar set-ups, for instance
from general relativity (no metrics with 8 or 9 Killing vectors in 4
dimensions), and supergravity (e.g.~no BPS solutions in eleven-dimensional
sugra preserving 31 supersymmetries \cite{Gran:2006cn}).

We focused on purely bosonic configurations, but solutions with non-trivial
fermionic fields are of course not excluded and are certainly worth studying.
Also, it would be desirable to work out explicit solutions to these various
sets of equations and verify whether their corresponding on-shell action is
finite. In the affirmative, the corresponding configurations would represent instantons possibly including non-trivial scalar and fermion fields profiles, which as of today are not known in closed form\footnote{Note that it has been argued that such a solution may not exist for generic gauge group \cite{Dorey:2002ik}}. The interest in instanton effects in ${\cal N}=4$ SYM is at least twofold. On the one hand, the theory is believed to be self-dual \cite{Montonen:1977sn}, a statement entailing the complete effective action including all instanton and anti-instanton effects. On the other hand, instantons have provided some of the most striking tests of the AdS/CFT correspondence. From this point of view, instantons with topological charge $k$ in ${\cal N}=4$ SYM with $SU(N)$ gauge group are obtained by adding $k$ D-instantons (D(-1) branes) to the stack of $N$ $D3$ branes \cite{Witten:1995im, Douglas:1996uz, Douglas:1995bn, Billo:2002hm} (for unoriented D-instantons, one starts with D3-branes on top of an orientifold 3-plane, the gauge group of the ${\cal N}=4$ SYM theory becoming $Sp(N)$ or $SO(N)$ depending on the charge of the O3-plane \cite{Hollowood:1999ev}). In the low energy supergravity limit where computations can mostly be performed (see however \cite{Bianchi:2007ft}, Sect. 18.3 for a discussion beyond sugra, in the BMN limit), D-instantons arise as non-trivial solutions of the Euclidean field equations. The classical type IIB supergravity action in $AdS_5 \times S^5$ can take these into account by incorporating the effect of the infinite tower of massive string excitations on the dynamics of the massless modes. In the case of minimal correlators/$AdS$ amplitudes, it turns out there is a perfect agreement between instanton contributions to SYM correlation functions and D-instanton induced supergravity amplitudes (see \cite{Bianchi:1998nk, Dorey:1998xe, Dorey:1999pd}, or \cite{Bianchi:2007ft}, Sects.15-18, \cite{Belitsky:2000ws} for reviews). Of course, this agreement is striking since the computations on the field theory side are done at weak coupling, and indicates that the corresponding correlators are protected from quantum corrections. On the gauge theory side, the latter computations are performed in a particular instanton background. The latter is generated from the self-dual configuration, starting from the YM instanton and solving iteratively the full set of coupled equations to get an approximated truncated solution by retaining terms only up to a certain power of the coupling constant, which is enough to compute correlators in the semi-classical approximation, see e.g. Sect.14 of \cite{Bianchi:2007ft}).
If new finite-action solutions would appear to exist, in particular with non (anti-)selfdual field strength, it would certainly be a very interesting problem to compute their contribution to correlation functions and match it with a dual supergravity computation.

Finally, the AdS/CFT correspondence has also allowed to give a string theory
interpretation of SYM states.  For example, it is known that monopoles and
dyons are dual to D-strings and bound states of D-strings and fundamental
strings, respectively,  between different D3-branes. In the same spirit, 1/4
BPS states in ${\cal N}=4$ SYM with a gauge group SU(3) have been shown to
correspond to three-pronged strings connecting three
D3-branes \cite{Bergman:1997yw}.
It would therefore be really interesting to identify to which configurations on
the gravity side these various supersymmetric solutions (or their counterparts
in the lorentzian theory) are mapped through the AdS/CFT correspondence.
The strategy of the present analysis is of course not limited to four-dimensional ${\cal N}=4$ SYM, and could be applied to any supersymmetric gauge theory. An example of particular interest are the superconformal three-dimensional Chern-Simons theories recently proposed in the context of the $AdS_4/CFT_3$ corespondence \cite{Aharony:2008ug}. We hope to return to these questions in future works.

\acknowledgments

This work was partially supported by INFN, MIUR-PRIN contract 20075ATT78 and
by the European Community FP6 program MRTN-CT-2004-005104. We would like to
thank Sophie de Buyl, Patrick Meessen, Christoffer Peterson, Christoph Sieg, Kirsten Vogeler and
Vincent Wens for useful discussions. The work of S. D.~has been funded by INFN
and by the European Commission through the grant PIOF-GA-2008-219950 (Home
institution: Universit\'e Libre de Bruxelles, Service de Physique Th\'eorique et
Math\'ematique, Campus de la Plaine, B-1050 Bruxelles, Belgium).

\normalsize

\appendix

\section{Conventions}
\label{conv}

In this appendix, we collect some of our conventions and notations in relation
to spinors and Clifford algebras based on \cite{Vandoren:2008xg}.

The 't Hooft symbols are defined as
\begin{eqnarray}
\eta_{a\mu\nu}&=&\epsilon_{a\mu\nu}+\delta_{a\mu}\delta_{\nu 4}-\delta_{a\nu}
                \delta_{4\mu}\,, \nonumber \\
\bar\eta_{a\mu\nu}&=&\epsilon_{a\mu\nu}-\delta_{a\mu}\delta_{\nu 4}+\delta_{a\nu}
                    \delta_{4\mu}\,, \label{eq: hooft}
\end{eqnarray}
with $a=1,2,3$ and $\mu,\nu=1,2,3,4$. The three matrices $\eta_a$
are selfdual, while the  $\bar\eta_a$ are anti-selfdual,
\begin{equation}
\frac12{\epsilon_{\mu\nu}}^{\rho\sigma}\eta_{a\rho\sigma}=\eta_{a\mu\nu}\,, \qquad
\frac12{\epsilon_{\mu\nu}}^{\rho\sigma}\bar\eta_{a\rho\sigma}=-\bar\eta_{a\mu\nu}\,,
\qquad (\epsilon_{1234}=1)\,,
\end{equation}
and together they form a basis for the $4\times 4$ antisymmetric matrices.
Moreover, they satisfy the relations
\begin{eqnarray}
&&[\eta_a,\eta_b]=-2\epsilon_{abc}\eta_c\,, \qquad
  [\bar\eta_a,\bar\eta_b]=-2\epsilon_{abc}\bar\eta_c\,, \nonumber \\
&&\{\eta_a,\eta_b\}=-2\delta_{ab}\,, \qquad \{\bar\eta_a,\bar\eta_b\}=
  -2\delta_{ab}\,, \\
&&\left[\eta_a,\bar\eta_b\right]=0\,. \nonumber
\end{eqnarray}
When discussing spinors in 6 dimensions, we will use the notation 
\begin{equation}
  \vec{\eta}=(\eta_1,\eta_2,\eta_3)\,, \qquad
  \vec{\bar\eta}=(\bar\eta_1,\bar\eta_2,\bar\eta_3)\,,
\end{equation}
understood as three-component vectors of $4\times 4$ matrices.

We use the following representation of the four-dimensional euclidean Clifford
algebra:
\begin{equation}
  \gamma_\mu=
  \left(
  \begin{array}{cc}
    0 & -i\sigma_\mu\\
    i\bar\sigma_\mu & 0
  \end{array}
  \right)\,, \label{rep-gamma}
\end{equation}
with
\begin{equation}
\sigma_\mu=(\vec\tau,i)\,, \qquad \bar\sigma_\mu=(\vec\tau,-i)\,, \qquad
\mu=1,\ldots,4\,,
\end{equation}
$\vec\tau$ denoting the three Pauli matrices. In this representation the
Spin$(4)$ generators on four-component Dirac spinors are
\begin{eqnarray}
&&\gamma_{\mu\nu}=
    \left(
    \begin{array}{cc}
      \sigma_{\mu\nu} & 0\\
      0 & \bar\sigma_{\mu\nu}
    \end{array}
    \right)\,, \label{gamma_munu} \\
&&\sigma_{\mu\nu}=\frac12\left(\sigma_\mu\bar\sigma_\nu-\sigma_\nu\bar\sigma_\mu
\right)=i\bar\eta_{a\mu\nu}\tau^a\,, \label{sigma} \\
&&\bar\sigma_{\mu\nu}=\frac12\left(\bar\sigma_\mu\sigma_\nu-\bar\sigma_\nu
\sigma_\mu\right)=i\eta_{a\mu\nu}\tau^a\,, \label{barsigma}
\end{eqnarray}
where the relationship to the 't Hooft symbols has been made explicit. The
matrices $\sigma_{\mu\nu}$ and $\bar\sigma_{\mu\nu}$ are the generators of the
two inequivalent pseudo-real irreducible representations of Spin$(4)$
acting on two-component Weyl spinors $\psi^\alpha$ and $\bar\chi_{\dot{\alpha}}$
respectively. Indices $\alpha,\dot{\alpha}=1,2$ are raised and lowered
according to the north-west convention
\begin{equation}
\epsilon^{\alpha\beta}\psi_\beta=\psi^\alpha\,, \qquad
\bar\psi^{\dot\beta}\epsilon_{\dot\beta\dot\alpha}=\bar\psi_{\dot\alpha}\,,
\end{equation}
where the antisymmetric invariant tensor $\epsilon$ is defined by
\begin{displaymath}
\epsilon_{12}=1\,, \qquad \epsilon^{\alpha\beta}=\epsilon_{\alpha\beta}\,, \qquad
\epsilon_{\beta\alpha}=-\epsilon_{\alpha\beta}\,, \qquad
\epsilon_{\dot\alpha\dot\beta}=\epsilon^{\dot\alpha\dot\beta}=
-\epsilon_{\alpha\beta}\,, \qquad
\epsilon_{\dot\beta\dot\alpha}=-\epsilon_{\dot\alpha\dot\beta}\,.
\end{displaymath}
In 6 dimensions, one defines the four by four matrices 
\begin{equation}
\Sigma^a=(-i\eta_1,\eta_2,\eta_3,i\vec{\bar\eta})\,, \qquad
\bar\Sigma^a=(i\eta_1,-\eta_2,-\eta_3,i\vec{\bar\eta})\,,
\end{equation}
with elements $\Sigma^{a,AB}$ and $\bar\Sigma^a_{AB}$, where indices
$a=1,\ldots,6$ are raised with the flat Minkowski metric in 5+1 dimensions.
The matrices
\begin{equation}
  \hat\gamma_a=
  \left(
  \begin{array}{cc}
    0 & \Sigma_a\\
    \bar\Sigma_a & 0
  \end{array}
  \right)
\end{equation}
form a representation of the Clifford algebra $Cl(5,1)$. The corresponding
representation of Spin$(5,1)$ is again reducible into two pseudo-real
inequivalent representations, with right Weyl spinors transforming with
$\Sigma_{ab}=\frac12(\Sigma_a\bar\Sigma_b-\Sigma_b\bar\Sigma_a)$ and left Weyl
spinors with
$\bar\Sigma_{ab}=\frac12(\bar\Sigma_a\Sigma_b-\bar\Sigma_b\Sigma_a)$, i.e.,
\begin{equation}
{\hat\gamma}_{ab} = \left(\begin{array}{cc}
                    \Sigma_{ab} & 0 \\ 0 & {\bar\Sigma}_{ab}
                    \end{array}\right)\,. \label{hatgamma_ab}
\end{equation}

\end{document}